\documentclass[letterpaper,11pt,openany]{article}
\usepackage{fullpage}
\usepackage{enumerate}
\usepackage[T1]{fontenc}
\usepackage[utf8]{inputenc}
\usepackage{natbib}
\usepackage{comment}
\usepackage{float}
\usepackage{hyperref}
\usepackage{mathrsfs}
\usepackage{enumitem}
\usepackage[font={small,it}]{caption}
\usepackage{amsmath,amsfonts,amsthm,amssymb}
\usepackage{bm,rotating,multirow,dsfont,graphicx}
\usepackage[usenames, dvipsnames]{color}
\usepackage{url}
\usepackage{multicol}
\usepackage{multirow}
\usepackage[T1]{fontenc}
\usepackage{flafter}
\usepackage{appendix}
\usepackage{subfigure}
\usepackage{xcolor}
\hypersetup{
	colorlinks,
	linkcolor={black},
	citecolor={black},
	urlcolor ={black}
}

\makeatletter
\def\hlinewd#1{%
\noalign{\ifnum0=`}\fi\hrule \@height #1 %
\futurelet\reserved@a\@xhline}
\makeatother
\newcommand\simiid{\mathrel{\overset{\makebox[0pt]{\mbox{\normalfont\tiny\sffamily IID}}}{\sim}}}
\newcommand\simind{\mathrel{\overset{\makebox[0pt]{\mbox{\normalfont\tiny\sffamily IND}}}{\sim}}}
\newcommand{\expo}[1]{\exp{ \left\{ #1 \right\}}}
\newcommand{\bs}{\boldsymbol}

\newcommand{\al}{\alpha}
\newcommand{\be}{\beta}
\newcommand{\sig}{\sigma}

\newcommand{\yv}{\boldsymbol{y}}
\newcommand{\xv}{\boldsymbol{x}}
\newcommand{\Xv}{\boldsymbol{X}}
\newcommand{\bev}{\boldsymbol{\beta}}
\newcommand{\tev}{\boldsymbol{\theta}}
\newcommand{\muv}{\boldsymbol{\mu}}

\newcommand{\Sig}{\mathbf{\Sigma}}
\newcommand{\SIG}{\mathbf{\Sigma}}
\newcommand{\LAM}{\mathbf{\Lambda}}
\newcommand{\Sm}{\mathbf{S}}
\newcommand{\Xm}{\mathbf{X}}
\newcommand{\I}{\mathbf{I}}

\def\normal{\small{\mathsf{N}}}
\def\GI{\small{\mathsf{GI}}}
\def\trans{\small{\mathsf{T}}}
\allowdisplaybreaks
\begin{document}

\renewcommand{\tablename}{Tabla}

\title{Bayesian Analysis for Social Science Research}

\author{
Carolina Luque, Universidad Ean, Colombia\footnote{cluque2.d@universidadean.edu.co} \\
Juan Sosa, Universidad Nacional de Colombia, Colombia\footnote{jcsosam@unal.edu.co}
}

\date{}

\maketitle

\begin{abstract} 
In this manuscript, we discuss the substantial importance of Bayesian reasoning in Social Science research.
Particularly, we focus on foundational elements to fit models under the Bayesian paradigm. 
We aim to offer a frame of reference for a broad audience, not necessarily with specialized knowledge in Bayesian statistics, yet having interest in incorporating this kind of methods in studying social phenomena. 
We illustrate Bayesian methods through case studies regarding political surveys, population dynamics, and standardized educational testing.
Specifically, we provide technical details on specific topics such as conjugate and non-conjugate modeling, hierarchical modeling, Bayesian computation, goodness of fit, and model testing.

\end{abstract}


\newpage

\section{Introduction}

Bayesian methods refer to data analysis tools derived from the principles of Bayesian inference, i.e., inductive learning through Bayes' Theorem \citep{van2021bayesian}. 
These methods allow us to estimate parameters with good statistical properties, predicting/imputing future/missing observations, and making optimal decisions according to prespecified utility functions. 
Moreover, Bayesian techniques are also intrinsically linked to sophisticated computational algorithms for model estimation, selection, and validation \citep{hoff2009first}. 
A wide variety of scientific publications in the context of Social Sciences point out that Bayesian methods are a predominant methodology for the analysis of different phenomena since the early 1990s \citep[e.g.,][]{western1994bayesian,jackman2004bayesian,gill2005elicited,barbera2015birds,moser2019multiple, lynch2019bayesian, fairfield2022social}.

Several reasons justify adopting a Bayesian approach to statistical inference in Social Sciences. 
Quantitative research shows that the social phenomenon is quite different from its counterpart in the experimental sciences, so its characteristics and methodological requirements are more akin to the Bayesian paradigm, far from the assumptions of the frequentist approach \citep[e.g.,][]{jackman2009bayesian, fairfield2019dialogue}. 
That is why we discuss advantages (and challenges!) about adopting a Bayesian spirit in Social Science research.

We provide foundations about model fitting under the Bayesian paradigm, from the prior distribution and sampling distribution specification to posterior inference mechanisms, including model verification and evaluation. 
Additionally, we illustrate the essentials of Bayesian methodologies through case studies in contexts such as political surveys, population dynamics, and standardized educational testing.

Unlike other authors \citep{lenhard2022transformation,sosabuitrago2022,van2021bayesian,kruschke2021bayesian,van2014gentle,draper2009bayesian, walker2007application, jackman2004bayesian}, we provide a review of the Bayesian paradigm focused exclusively on Bayesian machinery framed in Social Science research. 
We aim to offer a frame of reference for a broad audience, not necessarily with specialized knowledge in Bayesian statistics, yet having interest in incorporating this kind of methods in studying social phenomena.

This paper is structured as follows. Section \ref{sec:FB} draws a parallel between Bayesian and frequentist inference. Section \ref{sec:sociales} presents the rationale for a Bayesian approach to scientific research in Social Science. Section \ref{sec:fund} shows the importance of sensitivity analysis, the suitability of conjugate analysis, model evaluation and testing, and Bayesian computation via Monte Carlo simulation. Section \ref{sec:ilustracion} provides fully develops case studies from a Bayesian perspective. Finally, Section \ref{sec:discusion} discusses our main findings as well as some alternatives for future research.

\section{Statistical inference: frequentist versus Bayesian} \label{sec:FB}

Two paradigms to data analysis coexist in Statistics: \textit{frequentist} and \textit{Bayesian}. They differ in many ways, including the probability interpretation, the parameters' nature, and the statistical/computational methods required to make inferences. However, both alternatives are governed by axiomatic foundations of probability and use the likelihood function to estimate unknown parameters.

Under any of these approaches, the relationship between data $\boldsymbol{y}=(y_1,...,y_n)$ and parameters $\theta\in\Theta$ is established by a sampling distribution $\boldsymbol{y} \sim p(\boldsymbol{y}\mid \theta)$, which fully characterizes the random mechanism that generates $\boldsymbol{y}$, for any given value of $\theta$. 
The nature of $\theta$ is one of the main differences between the Bayesian and the frequentist paradigm. On the one hand, the frequentist approach assumes $\theta$ as a fixed but unknown quantity, and any estimate of it constitutes a random variable in itself, since depends on repeated random sampling.
Typically, $\theta$ is estimated by maximizing $p(\boldsymbol{y}\mid \theta)$ as a function of $\theta$.
On the other hand, the Bayesian approach assumes $\theta$ as a random variable so that any estimate of it, is fixed and constitutes a realization of such a random quantity. 
From this point of view, we can directly incorporate into the model any prior beliefs (state of knowledge) about $\theta$ using probabilistic statements. 
Such a task can be carried out through the \textit{prior distribution}, $p(\theta)$, whose purpose is to characterize the uncertainty about $\theta$ external to $\boldsymbol{y}$. 
Thus, once $\boldsymbol{y}$ is observed, prior beliefs are updated, and the \textit{posterior distribution}, $p(\theta\mid\boldsymbol{y})$, is obtained in order to fully describe the updated state of knowledge about $\theta$, given the empirical evidence provided in $\boldsymbol{y}$. 
In this context, Bayes' Theorem is the optimal rational method that guarantees coherence and logical consistency for updating prior beliefs about $\theta$ according to the information contained in $\boldsymbol{y}$ \citep{jackman2004bayesian,hoff2009first}.

Bayes' Theorem states that 
\begin{equation*}
p(\theta\mid\boldsymbol{y})=
\frac{p(\boldsymbol{y} \mid \theta)\,p(\theta)}{\int_{\Theta} p(\boldsymbol{y} \mid \theta)\,p(\theta) \, \text{d}\theta}\,,
\end{equation*}
where $p(\boldsymbol{y})=\int_{\Theta} p(\boldsymbol{y} \mid\theta)\,p(\theta)\, \text{d}\theta$ is the marginal probability of $\boldsymbol{y}$, which does not depend on $\theta$ because it is an integral over all the values of $\theta\in\Theta$, and therefore, corresponds to a normalization constant that allows $p(\theta\mid\boldsymbol{y})$ to be a valid probability distribution. 
Since $p(\theta\mid\boldsymbol{y})$ and $p(\theta)$ are functions of $\theta$, and $p(\boldsymbol{y})$ does not depend on $\theta$, the sampling distribution $p(\boldsymbol{y}\mid \theta)$ has to be regarded as a function of $\theta$ (as frequentists do!), and therefore, Bayes' Theorem can be expressed as
$p(\theta\mid \boldsymbol{y}) \propto \ell(\theta) \,p(\theta)$,
where $\ell(\theta) = \text{c}\,p(\boldsymbol{y}\mid \theta)$ is the so called  \textit{likelihood function}, for any $\text{c} > 0$ (typically chosen as $\text{c} = 1$).

The previous expression makes evident two important aspects: (i) the posterior distribution is simply proportional to the likelihood function times the prior distribution, and (ii) some frequentist results can be seen as a particular cases of Bayesian analysis. 
Regarding the first aspect, the influence of prior beliefs and data on the posterior distribution depends on the amount of information provided in the prior distribution and the sample size, respectively. 
Regarding the second aspect, frequentist and Bayesian are typically equivalent when the prior distribution is non-informative (all the possible parameter values have the same density) and/or the sample size is large in comparison with the dimension of the parameter space. 
In this case, the posterior distribution takes the same form as the likelihood function.

Generally, Bayesian analysis are simple and direct from a conceptual point of view since they mainly rely on a naive application of Bayes' Theorem. 
\cite{cox1946probability}, \cite{cox1963algebra} and \cite{savage1972foundations} constitute theoretical support to justify that if $p(\boldsymbol{y} \mid \mathbf{\theta})$ and $p(\mathbf{\theta})$ represent the beliefs of a rational person (in a probabilistic sense), then Bayes' Theorem is the optimal method for updating his beliefs according to the laws of probability.
Furthermore, uncertainty quantification under the Bayesian paradigm is for free!
Once the posterior distribution is obtained, any aspect of the model parameters can be described using probabilistic statements. 
However, everything is not bright and there are challenges. Computation under the Bayesian paradigm is quite challenging because in multiparameter models integrals to characterize any aspect of the posterior distribution can be hard to compute.

\section{Bayesian inference in Social Science}\label{sec:sociales}

In social research, frequentist methods become unreliable when data do not correspond to a random sample from a larger population.
The conventional interpretation of confidence intervals as well as hypothesis tests, which rely on patterns emerging from repeated sampling, tend to be confusing and inadequate in contexts where uncertainty do not arise from variation in repeated sampling \citep{western1994bayesian,jackman2009bayesian}. 
Consequently, adhering to a frequentist notion of probability in the absence of repeatable data loses its meaning. 
In contrast, the subjective interpretation of probability provided in the Bayesian paradigm offers a coherent and internally consistent tool for statistical inference when data can not be framed in the context of repeated experimentation.

Social scientists often encounter themselves working with ``small'' data gathered from real-life social behavior, where classical experimental-design requirements are typically not met. 
In such scenarios, subjective judgments regarding the model's formulation become inevitable and intrinsic to the scientific process \citep{jackman2009bayesian}. 
Thus, it is natural for researchers in this context setting prior probabilities about the unknown quantities and interpret them subjectively (depending on the modeler's state of knowledge). 
Furthermore, the absence of ``large datasets'' implies that estimates obtained through a frequentist approach lack robust statistical properties, particularly concerning the asymptotic properties that validate classical inference procedures. 
Comparative research studies in the social domain have demonstrated that applying frequentist methods to small data may lead to imprecise estimates of the effects of explanatory variables \citep[e.g.,][]{western1994bayesian}.

The social field abounds with data grouped over several units or periods. Hence, a key research question is how a causal structure operating at one level of analysis (e.g., individuals) varies over a higher level (e.g., localities or periods). 
The Bayesian approach to statistical inference is well suited to answer this question since it allows the researcher to formalize assumptions about  between and within between-group heterogeneity by formulation a proper structured of prior beliefs. 
Thus, the prior distribution, considered by critics of Bayesian inference as a weakness, provides a way to expand from a simple model to a model involving several sources of heterogeneity, which allows modelers analyzing cases where social research requires understanding the relative weight of unknown quantities at different levels \citep{jackman2009bayesian}.

Other advantages of Bayesian inference, not exclusive of Social Science, include its conceptual simplicity. 
As mentioned previously, Bayesian inference does not require to consider hypothetical data as frequentists do when developing confidence intervals and hypothesis tests. 
This is because the posterior distribution directly represents the most up-to-date information about the parameter, given nothing more than the observed data. 
Such a straightforward nature inherent in the Bayesian paradigm is quite appealing to social scientists, making it the primary methodology adopted by quantitative researchers in this field \citep{jackman2009bayesian}.

Nowadays, Bayesian computation has become more feasible than ever before due to the current developments in both and hardware and software. 
Such a computational framework make possible to solve complex statistical problems that a few decades before were just not possible to handle. 
Specifically, the recent low cost and computational speed make it attainable for social scientists to analyze data from a Bayesian simulation-based approach. 
In this sense, the set of algorithms known as Markov Chain Monte Carlo (MCMC, see Sec. \ref{sec:MCMC} for details) allows the Bayesian approach to be a practical reality for applied researchers \citep{jackman2009bayesian}.
These algorithms provide a powerful and flexible way to approximate the posterior distribution for most multiparameter models. For example, estimates of hierarchical models, latent variable models, and estimates based on observations with missing data have become straightforward procedures because the mathematics and computation underlying Bayesian analysis is drastically simplified via Monte Carlo simulation.

\section{Fundamentals for Bayesian modeling}\label{sec:fund}

Here, we provide essential details for formulating and fitting Bayesian models, from conjugate families to specifics in Bayesian computation based on Monte Carlo simulation. This material is key to understand and implement the case studies given in Sec. \ref{sec:ilustracion}.

\subsection{Sensitivity analysis}

The main criticism of the Bayesian approach relies on the inherent subjectivism of the prior distribution.
For frequentists, data analysis based on subjective information states (depending on the analyst) lacks scientific rigour. 
However, the question is, why should we neglect available external information that is consistent with reality when it can contribute to explaining the phenomenon of interest and lead to more accurate inferences and plausible conclusions? 
Even when external information is not available, we can set the prior distribution to reflect such state of knowledge. In this regard, there are available a number of options to specify the prior distribution in an ``objective'' fashion thought the so called objective priors (see \citealt[Chap. 2]{reich2019bayesian} for details).

Since the state of information may vary depending on the analyst, the choice of the prior distribution and the robustness of the inferences based on this choice is a fundamental issue in Bayesian inference. Regarding the prior choice, the literature exposes some methods for eliciting subjective prior distributions \citep[e.g.,][]{berger2013statistical,congdon2007bayesian,garthwaite2005statistical,stuart1994kendall}. However, it may not be easy to formulate the researcher's prior beliefs mathematically and precisely. For this reason, the prior distribution is often only an approximation of such beliefs and can be chosen for computational convenience \citep{hoff2009first}.

From a Bayesian perspective, studying the robustness of inferences means performing a \textit{sensitivity analysis}. The purpose is to examine how the posterior distribution changes as different values of the hyperparameters are adopted. This analysis allows us to argue that the conclusions are consistent from both a qualitative and quantitative point of view. There are two ways to perform a sensitivity analysis in practice, either (i) by weakening or strengthening the adopted prior or (ii) by repeating the analysis with different priors \citep{jackman2004bayesian}.

\subsection{Conjugate distributions}\label{sec:conjugadas}

Conjugate families are essential in Bayesian statistics because they greatly simplify computation. 
Specifically, suppose that the prior distribution $p(\theta)$ belongs to a known family of distributions. 
Then, such a prior is said to be conjugate regarding the sampling distribution $p(\yv \mid \theta)$, if the posterior distribution $p(\theta \mid \yv)$ belongs to the same family of distributions as the prior does \citep{jackman2004bayesian, jackman2009bayesian}.
It is up to modelers to work them or not depending on their prior beliefs and their modeling choices.

Under conjugacy, the update from the prior to the posterior distribution merely changes the parameters that define the corresponding conjugate family. This feature is easy to interpret, and besides easing computation, this characteristic allows us to develop some intuition about Bayesian learning through straightforward examples. 
However, it's important to acknowledge that conjugate priors have certain limitations. For instance, not all likelihood functions have a known conjugate prior, and most conjugacy pairs are applicable only to small-scale examples with a limited number of parameters \citep{reich2019bayesian}. Furthermore, not every state of knowledge about an unknown parameter is easy to express using a conjugate prior distribution.

\subsection{Bayesian Computation via Monte Carlo Simulation} \label{sec:MCMC}

In Bayesian inference, we can formulate complex conjugate analyses, from a mathematical point of view, and even non-conjugate analyses, which motivate alternative methods to explore the posterior distribution. 
These methods allow the generation of random samples from the posterior distribution, in some cases computationally intensive, when the underlying computations required for the analysis are either extremely demanding or analytically intractable, typically due to the number of model parameters.
Simulation-based procedures to approximate the posterior distribution as well as other distributions or measures related to it are based on the \textit{Monte Carlo principle} and \textit{Markov chains} \citep[e.g.,][]{gamerman2006markov,robert2013monte,turkman2019computational}.

The Monte Carlo principle states that any characteristic of a random variable, can be approximated arbitrarily well by generating random samples from its probabilistic distribution. 
The learning accuracy rate strongly depends on the number of samples gathered during the simulation process. 
On the other hand, Markov chains are a first-order stochastic process, which allows exploring multidimensional parameter spaces when the posterior distribution has a complex or unknown analytic shape. 
Through this process, we are able to generate a random sequence (also known as random walk) of values with serial dependency, in order to empirically explore the probability distribution of a given random variable, even when the samples are correlated (for formal treatment of Markov chain theory see \citealp{carlin2008bayesian}, \citealp{jackman2009bayesian}, \citealp{meyn2012markov}, and \citealp{norris1998markov}).
In particular, \citealp{bradley2014matrices} provide applications in Social Sciences.

Monte Carlo samples are not a mechanism for gathering ``new data'' but rather to generate either independent and identically distributed (IID) or correlated samples from the posterior. In other words, these samples are a means to explore the posterior distribution, not an inference method in themselves.
Particularly, Markov chain theory states that it is possible to construct random walks in parameter spaces such that the chain visits locations in that space with frequencies proportional to the probability of those locations under a distribution of interest such as the posterior distribution. A chain with this property is called \textit{ergodic} \citep{jackman2004bayesian}. Ergodicity is an essential property because it ensures that we can simulate Markov chains whose stationary distribution is the posterior distribution, when the number of iterations grows.

\newpage

In practice, the iterative history generated by the Markov chain is typically stored and treated as a series of random samples from the posterior distribution. 
Then, following the Monte Carlo principle, these samples are useful for learning about any aspect of $\theta$ according to its posterior distribution. 
Thus, both the Monte Carlo principle and Markov chain theory support a whole range of \textit{Markov Chain Monte Carlo} (MCMC) algorithms, such as the \textit{Metropolis-Hastings algorithm} (Sec. \ref{sec:Metropolis}), the \textit{sampler Gibbs algorithm} (Sec. \ref{sec:Gibbs}) and the \textit{Hamiltonian Monte Carlo algorithm} (Sec. \ref{sec:Hamiltoniano}), among many other, which nowadays are core computational tools to carry out Bayesian inference.
For a review of modern methods in this regard, see part III of \cite{gelman2013bayesian}.

In Bayesian inference, Monte Carlo methods based on IID sampling an MCMC algorithms serve to the same purpose: Generating random sequences of $\theta$ that allows to fully characterize its posterior distribution. 
In the case of IID sampling, such independent samples constitute a direct representation of the posterior distribution, which can not be guaranteed for MCMC samples. 
Given the serial dependency of the samples that arise when implementing a MCMC algorithm, it is simply not possible to guarantee that they correspond to an empirical distribution close to the posterior distribution for a finite number of iterations. In other words, given the finite nature of the MCMC algorithm along with the autocorrelation between consecutive values of the random walk, there is no certainty at all that the simulated chain has reached convergence to the target distribution.

Evaluating convergence and precision with which Markov chains approximate the posterior distribution is not straightforward. 
In practice, it is common to diagnose non-convergence through graphical representation of the chain (e.g., traceplots) and numerical measures (e.g., $\hat{\text{R}}$ statistic, \citealt{gelman2013bayesian}). 
Although several authors have proposed strategies to examine convergence \citep[e.g.,][]{gelman1992inference,geweke1991evaluating,raftery1991many}, such approaches have highlighted the problem of establishing such diagnoses and have not guaranteed an infallible method to evaluate it \citep {geyer1992practical}. Consequently, in order to reach a reasonable approximation of the target distributio, using MCMC algorithms, it is highly recommended to run the algorithm a large number of times to ensure convergence. In particular, a higher number of iterations compared to what would be need using IID sampling \citep{hoff2009first}.

Aiming to increase as much as possible the \textit{effective sample size} (equivalent sample size under IID sampling)  of a sequence of samples generated using MCMC algorithms, it is customary to discard the initial values of the chain (burn-in period) and take systematic samples of it (thinning). 
It is also highly recommended to run several chains at different starting points of the parameter space to check whether they tend to the same stationary distribution or not. 
During this process, it may be possible to encounter processing or storage limitations, depending on the number of models parameters and the complexity of the model. Therefore, highly optimized algorithms are in order.

\subsubsection{Monte Carlo principle}\label{sec:MonteCarlo}

Let $\theta$ be the parameter of interest and $\yv$ be a sample of observed values from a posterior distribution $p(\theta \mid \yv)$. Suppose that a IID random sample of size $B$ is drawn from $p(\theta \mid \yv)$, i.e.,
$\theta^{(1)}, \cdots, \theta^{(B)} \stackrel{\text{i.i.d}}{\sim} p(\theta \mid \yv)$.
Thus, the empirical distribution induced by $\theta^{(1)}, \cdots, \theta^{(B)}$ is known as the Monte Carlo approximation of the target distribution $p(\theta \mid \yv)$. 
Such empirical distribution gets closer to the true target distribution as $B$ increases. 
In practice, it is customary to choose $B$ large enough such that the Monte Carlo standard error (i.e., the standard deviation of the Monte Carlo samples divided by the square root of $B$) is less than the desired precision \citep{hoff2009first}. 
Additionally, thanks to the law of large numbers states, we have that
$$\frac{1}{B}\sum_{b=1}^{B}g(\theta^{(b)}) \longrightarrow \mathsf{E}(g(\theta)\mid \yv) = \int_{\Theta} g(\theta)\,p(\theta \mid \yv)\, \text{d}\theta\, \hspace{0.2cm} \text{as long as} \hspace{0.2cm} B \rightarrow\infty\,,
$$
where $g(\theta)$ is any function of $\theta$.
Consequently, any aspect of the posterior distribution can be approximated arbitrarily well with a large enough Monte Carlo sample \cite{hoff2009first}.

\subsubsection{Gibbs sampler}\label{sec:Gibbs}

When it is difficult to simulate from the posterior distribution directly, it is recommended to sample iteratively from the \textit{full conditional distribution} $p(\theta_{i} \mid \theta_ {1}, \ldots, \theta_{i-1},\theta_{i+1}, \ldots, \theta_{k}, \yv)$ of each parameter $\theta_i$, for $i =1,\cdots,k$. 
The Gibbs sampler allow us to generate samples from the posterior distribution by updating sequentially each component of $\theta$ through its conditional distribution, given the most recent state of the other model parameters.
Specifically, given the current state of the parameters $\theta^{(b-1)}=(\theta_{1}^{(b-1)},\ldots,\theta_{k}^{( b-1)})$, we can generate the next state $\theta^{(b)}$ from $\theta^{(b-1)}$, for $b=1,\ldots,B$, as follows:
\begin{enumerate}
\item[$1$.]  Draw  $\theta_1^{(b)}  \sim  p(\theta_{1} \mid \theta_{2}^{(b-1)},\theta_{3}^{(b-1)}, \ldots, \theta_{k}^{(b-1)})$.
\item[$2$.]  Draw  $\theta_2^{(b)}  \sim  p(\theta_{2} \mid \theta_{1}^{(b)},  \theta_{3}^{(b-1)}, \ldots, \theta_{k}^{(b-1)})$.
\item[$ $] \hspace{1.8cm}$\vdots$
\item[$k$.]  Draw  $\theta_k^{(b)}  \sim  p(\theta_{k} \mid \theta_{1}^{(b)},  \theta_{2}^{(b)},   \ldots, \theta_{k-1}^{(b)})$.
\end{enumerate}
This algorithm generates a dependent sequence of values of $\theta$, namely, $\theta^{(1)},\ldots,\theta^{(B)}$.
In this random sequence, $\theta^{(b)}$ depends on $\theta^{(0)},\theta^{(1)}\ldots,\theta^{(b-1)}$ only through $\theta^{(b-1)}$, which means that, given $\theta^{(b-1)}$, $\theta^{(b)}$ is conditionally independent of $\theta^{(0)},\theta^{(1)}\ldots,\theta^{(b-2)}$ (this is the so called the Markov property).
Finally, the target distribution is reached as $b\longrightarrow\infty$, no matter what starting value $\theta^{(0)}$ is chosen to start the algorithm (although some starting are more convenient than others).

\subsubsection{Metropolis-Hastings}\label{sec:Metropolis}

Again, when it is difficult or even possible to simulate from the posterior distribution directly, the Metropolis-Hastings algorithm provides a general setting to build a Markov chain through a series of ``jumps'' that generate a random sequence, whose target distribution is the posterior distribution $p(\theta \mid \yv)$.
Specifically, given the current state of the parameters $\theta^{(b-1)}=(\theta_{1}^{(b-1)},\ldots,\theta_{k}^{( b-1)})$, we can generate the next state $\theta^{(b)}$ from $\theta^{(b-1)}$, for $b=1,\ldots,B$, as follows:
\begin{enumerate}
    \item Simulate a jump candidate $\theta^*$ around $\theta^{(b-1)}$ using a proposal distribution $J(\theta^* \mid \theta^{(b-1)})$. Usually, $J(\theta^* \mid \theta^{(b-1)})$ is taken to be symmetric, i.e., $J(\theta^* \mid \theta^{(b-1)}) =  J(\theta^{(b-1)}\mid \theta^*)$ (in this case, the algorithm is simply known as Metropolis algorithm). For instance, when $\theta$ is univariate, commonly used proposal distributions are $\textsf{N}(\theta^* \mid \theta^{(b-1)}, \delta)$ and $\textsf{U}(\theta^* \mid \theta^{(b-1)} - \delta,\theta^{(b-1)} + \delta)$, where the tunning parameter $\delta$ is chosen to allow the algorithm run efficiently. In practice, it is common to set $\delta$ in such a way that the proportion of effective jumps roughly lies between 20 and 50\%.
    \item Compute the acceptance ratio 
    $$
    r=\frac{p(\theta^* \mid \yv) / J(\theta^* \mid \theta^{(b-1)})}{p(\theta^{( b-1)} \mid \yv) / J(\theta^{(b-1)} \mid \theta^*)}\,.
    $$ 
    If the proposal distribution is symmetric, then the acceptance rate becomes
    $$
    r=\frac{p(\theta^* \mid \yv)}{p(\theta^{( b-1)} \mid \yv)}\,.
    $$ 
    Typically, $r$ is expressed on logarithmic scale in order to achieve numerical stability.
    \item Determine the transition probability $\alpha = \min\{1,r\}$. Thus, if the candidate increases the probability of the posterior distribution, then it is accepted with probability 1. On the other hand, if the candidate does not increase the probability of the posterior distribution, then it is accepted with probability $r$. 
    \item Simulate $u \sim \textsf{U}(0,1)$.
    \item Set $\theta^{(b)}=\theta^*$, if $u \leq \alpha$, and $\theta^{(b)}=\theta^{(b-1)}$, otherwise.
\end{enumerate}
Again, it can be shown the algorithm given above, regardless of the proposal distribution $J(\cdot\mid\cdot)$ and the initial value $\theta^{(0)}$, generates a Markov chain whose stationary distribution is the posterior distribution $p(\theta \mid \yv)$ \citep{gamerman2006markov}. 
See also our discussion about the Markov property given in the previous section.

\subsubsection{Monte Carlo Hamiltonian}\label{sec:Hamiltoniano}

A possible inefficiency of the Gibbs sampler and the Metropolis-Hastings algorithm lies in their local random walk behavior \citep{gelman2013bayesian}, which causes the chain to take too long to explore the posterior distribution efficiently. Such a behavior leads to long-time converge times, mainly when dealing with complex models such as those related to high-dimensional posterior distributions \citep{betancourt2019convergence}.
The Hamiltonian Monte Carlo algorithm is an alternative to overcome such inefficiency.

This algorithm considers a ``boost'' variable $\varphi$ to explore more efficiently the target distribution by moving on different trajectories, suppressing the local random walk motion described by other samplers \citep{betancourt2017conceptual,betancourt2019convergence}. 
In a Hamiltonian algorithm, samples are drawn from the joint distribution $p(\theta, \varphi \mid \yv) = p(\theta \mid \yv)\,p(\varphi)$. 
However, only simulations of $\theta$ are of interest since $\varphi$ operates as an auxiliary variable.
Specifically, given the current state of the parameters $\theta^{(b-1)}=(\theta_{1}^{(b-1)},\ldots,\theta_{k}^{( b-1)})$, we can generate the next state $\theta^{(b)}$ from $\theta^{(b-1)}$, for $b=1,\ldots,B$, as follows:
\begin{enumerate}
    \item Simulate $\varphi \sim \textsf{N}(0, \mathbf{M})$, where $\mathbf{M}$ is a diagonal matrix representing the covariance matrix associated with the impulse function $p(\varphi )$. Typically, $\mathbf{M}$ is chosen to be as the identity matrix. 
    \item Update $(\theta, \varphi)$ using $L$ ``jumps'' scalded by a factor $\epsilon$. Specifically, in a given jump, both $\theta$ and $\varphi$ change relative to each other as follows:
    \begin{enumerate}
        \item Update $\varphi$:
        $$\varphi \leftarrow \varphi + \frac{\epsilon }{2}\,\frac{\partial}{\partial \theta}\log p(\theta \mid \yv)\,.$$
        \item Update $\theta$:
        $$\theta \leftarrow \theta + \epsilon\,\mathbf{M}\varphi\,.$$
        \item Repeat the above steps $L-1$ times.
    \end{enumerate}
    \item Let $\theta^{(b-1)}$ and $\varphi^{(b-1)}$ be the initial values of $\theta$ and $\varphi$ respectively, and $\theta^{*}$ and $\varphi^{*}$ the corresponding values after the $L$ steps. Compute the acceptance ratio 
    $$
    r= \frac{p(\theta^{*} \mid \yv)\,p(\varphi^{*})}{p(\theta^{(b-1)} \mid \yv )\,p(\varphi^{(b-1)})}\,.
    $$ 
     \item Determine the transition probability $\alpha = \min\{1,r\}$.
    \item Simulate $u \sim \textsf{U}(0,1)$.
    \item Set $\theta^{(b)}=\theta^*$, if $u \leq \alpha$, and $\theta^{(b)}=\theta^{(b-1)}$, otherwise.
\end{enumerate}
The tunning parameters $\epsilon$ and $L$ are chosen to allow the algorithm run efficiently. In practice, it is common to set them in such a way that the proportion of effective jumps roughly lies between 60 and 70\%. See \cite{gelman2013bayesian} for more details about the choice of $\epsilon$, $L$ and $\mathbf{M}$.

\subsection{Goodness of fit} \label{sec:ajuste}

After establishing the structure of the model and approximating the posterior distribution $p(\theta \mid \yv)$, it is convenient to evaluate the model's fit, aiming to detect misleading inferences due to a poor model fitting. 
Formally, the model's goodness of fit can be carried out through external validation tests, which consist in generating hypothetical replicas of the data, say $\yv^{\text{rep}}$, though the posterior predictive distribution, $p(\yv^{\text{rep}} \mid \yv) = \int_\Theta {p(\yv^{\text {rep}} \mid \theta)\,p(\theta \mid \yv)} \, \text{d}\theta\,$.
Then, such replicated data are directly compered with the observed data. If the model fits well to the data, then replicated data should present a similar behavior to the observed data.

Usually, the model and data discrepancy is examined through a set of \textit{test statistics} (e.g., measures of trend and variability), say $t(\yv)$. These quantities are used as metrics to compare the predictive simulations with their corresponding observed values. In addition, such quantities allow us to identify the relevant aspects of the data that are reasonably reproduced by the proposed model.
The lack of fit of the data concerning the posterior predictive distribution is measured by the \textit{posterior predictive $p$ value}, $\text{ppp} = \textsf{Pr}(t(\yv ^{\text{rep}}) > t(\yv) \mid \yv)$, which can be interpreted as the probability that the replicated data is more extreme than the observed data (in test statistics terms). Thus, the model fits well to the data regarding the test statistic $t(\yv)$ if and only if the corresponding ppp does not assume extreme values such as 0 or 1 \citep{gelman2013bayesian}.

\subsection{Model comparisson} \label{sec:comparar}

Information criteria allow us to evaluate and compare models through their predictive performance. Poluar alternatives include the Deviance Information Criterion \citep[DIC, see][]{gelman2013bayesian, spiegelhalter2002bayesian} and the Watanabe-Akaike Criterion \citep[WAIC, see][]{gelman2013bayesian, watanabe2013waic}.

The DIC is defined as
$$ \text{DIC}= -2\log p(\yv \mid \hat\theta_{\text{Bayes}})+2p_{\text{DIC}} \, ,$$
where $\hat\theta_{\text{Bayes}}= \textsf{E}(\theta \mid \yv) \approx \frac{1}{B}\sum_{b=1}^{B}\theta ^{(b)}$ is the posterior mean of $\theta$, and $p_{\text{DIC}}$ to the effective number of parameters,
$$
p_{\text{DIC}} = 2 \left[ \log p(\yv \mid \hat\theta_{\text{Bayes}})- \textsf{E}(\log p(\yv \mid \theta)\mid\yv) \right]
\approx 2 \left[ \log p(\yv \mid \hat\theta_{\text{Bayes}}) - \frac{1}{B}\sum_{b=1}^{B} \log p\left(\yv \mid \theta^{(b)}\right) \right]\,.
$$

On the other hand, the WAIC is defined as
$$
\text{WAIC} = -2 \text{lppd} + 2 p_{\text{WAIC}},
$$
where  
$$\text{lppd}=\log \prod_{i=1}^{n} p(y_i \mid \yv) = \sum_{i=1}^{m} \log \int_{\Theta} p(y_i \mid \theta)\, p(\theta \mid \yv)\, \text{d}\theta \approx \sum_{i=1}^{n} \log \left ( \frac{1}{B}\sum_{b=1}^{B}p(y_i \mid \theta^{(b)}) \right )$$ 
is the posterior predictive distribution in logarithmic scale, which summarizes the predictive ability of the model fitted to the data.
The corresponding effective number of parameters is given by
$$p_{\text{WAIC}} = 2 \sum_{i=1}^{n} \left[\log \left(\textsf{E}(p(y_i \mid \theta)\mid\yv)\right)-\textsf{E}\left(\log( p(y_i \mid \theta)\mid\yv)\right) \right] \, ,$$
which in practice can be calculated as
$$
p_{\text{WAIC}} \approx 2\sum_{i=1}^{n} \left [ \log\left ( \frac{1}{B}\sum_{b=1}^{B}p(y_i \mid \theta^{(b)}) \right )-\frac{1}{B}\sum_{b=1}^{B} \log p(y_i \mid \theta^{(b)})\right ]\,.
$$
When Comparing models, lower DIC and WAIC values imply higher predictive accuracy.

Although the DIC is widely used as a model selection tool, it has several disadvantages compared to the WAIC. 
Common criticisms include the penalty term, $p_{\text{DIC}}$, is not invariant to reparameterization; the DIC may not be consistent with identical replicates of the same experiment; the DIC is not based on a completely Bayesian predictive criterion \citep[see][, for more details]{spiegelhalter2014deviance}. 
The WAIC addresses many of these criticisms. 
In particular, The WAIC is invariant to reparameterizations, which makes it useful in the case of models with hierarchical structures, in which the number of parameters increases with the sample size \citep{spiegelhalter2014deviance}.

\section{Study cases} \label{sec:ilustracion}

This section illustrates the Bayesian methodologies described in previous sections with three case studies.
First, we exemplify the Monte Carlo principle using IID sampling in the context of a multinomial-Dirichlet model.
Then, we illustrate the Metropolis-Hastings algorithm, the Hamiltonian Monte Carlo algorithm, and goodness-of-fit methods in the context of a generalized linear model for count data.
Finally, we show the Gibbs sampler and information criteria metrics in the context of hierarchical linear regression models.
The interested reader may request the code to reproduce all the examples from any of the authors.

\subsection{Political survey: A Multinomial-Dirichlet model}

We implement a Multinomial-Dirichlet model to analyze the 2022 Colombian Presidential Consultations. This model allows us to estimate the population share of votes that each candidate will receive based on the data provided by a national pollster.
The independent media company Valora Analitik reported that ``after adding up the differences between the latest polls and the results given in Election Day, Invamer is the pollster that was closest in its predictions, followed by Guarumo and EcoAnalítica, and in third place, the CNC. The pollster furthest away from the results was Yanhaas, in fourth place''\footnote{\url{https://www.valoraanalitik.com/2022/03/14/ranking-encuestadoras-elecciones-marzo-colombia-2022/}}. 
Consequently, we use the Invamer results to illustrate the way  a Multinomial-Dirichlet model is implemented.
The Invamer survey was conducted at the end of February 2022\footnote{\url{https://es.scribd.com/document/562600199/Invamer-Marzo-2022}}. It involved the participation of 1504 men and women aged 18 and over, representing diverse socio-economic levels across the country, including urban and rural areas. This survey seeks to gather information about participants' preferences in the presidential consultations for Colombia's elections in 2022. In Table \ref{tab_resultados}, data correspondes to respondents who indicated their definite or probable vote for each party's consultation. It is important to note that this count does not include undecided voters.

\begin{table}[H]
\centering
{\footnotesize
\begin{tabular}{cccccc}
\hline
\multicolumn{5}{c}{Pacto Histórico} &  \\ \hline
G. Petro & F. Márquez & C. Romero & A. U. Guariyú & A. Saade & $n$ \\ \hline
322 & 56 & 24 & 7 & 1 & 410 \\ \hline
\multicolumn{5}{c}{Coalición Equipo por Colombia} &  \\ \hline
F. Gutiérrez & A. Char & E. Peñalosa & D. Barguil & A. Lizarazo & $n$ \\ \hline
51 & 43 & 33 & 27 & 22 & 176 \\ \hline
\multicolumn{5}{c}{Coalición Centro Esperanza} &  \\ \hline
S. Fajardo & J. M. Galán & C. Amaya & A. Gaviria & J. E. Robledo & $n$ \\ \hline
45 & 28 & 18 & 15 & 13 & 119 \\ \hline
\end{tabular}}
\caption{Invamer's survey results about party consultations in Colombia 2022.}
\label{tab_resultados}
\end{table}

Although Invamer uses a particular kind of random sampling without replacement, it is customary to consider such a sample as a simple random sample with replacement, given that the total sample size is very small compared to the size of the Universe. 
Under the conditions given above and given that our uncertainty about the responses of the 1504 people in the survey is interchangeable, a particular version of De Finetti's Theorem \citep[p. 176]{bernardo2009bayesian} guarantees that the only sampling distribution appropriate for data of this nature is the Multinomial distribution. Below we describe the modeling approach as well as the results in the context of the political landscape in Colombia.

\begin{figure}
	\centering
	\includegraphics[scale=.5]{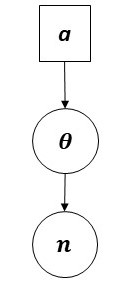}
	\caption{\footnotesize{DAG for the Multinomial-Dirichlet model.}}
	\label{fig_dag_multinomial}
\end{figure}

The population of interest consists of items categorized into $k\geq2$ types, where each type $j$ has a proportion denoted by $0<\theta_j<1$, with $j=1,\ldots,k$.
The components of $\boldsymbol{\theta}=(\theta_1,\ldots,\theta_k)$ are such that $\sum_{j=1}^k\theta_j=1$.
Now, an IID sample $\boldsymbol{y}=(y_1,\ldots,y_n)$ of size $n$ is taken from the population. 
Let $\boldsymbol{n}=(n_1,\ldots,n_k)$ be the random vector that represents the counts associated with each type of item. Here, $n_j$ denotes the number of elements in the random sample that belong to type $j$, for $j=1,\ldots,k$. 
In this situation, $\boldsymbol{n}$ follow a Multinomial distribution with parameters $n$ and $\boldsymbol{\theta}$, which is defined as follows: 
$\boldsymbol{n}\mid n,\boldsymbol{\theta}\sim\textsf{Mult}( n,\boldsymbol{\theta})$ if and only if 
\begin{equation}\label{eq_multinomial_1}
	p( \boldsymbol{n}\mid n, \boldsymbol{\theta} ) =  \frac{n!}{\textstyle\prod_{j=1}^k n_j!} \prod_{j=1}^k \theta_j^{ n_j }  
\end{equation}
provided that $\sum_{j=1}^k n_j = n$ and $0 \leq n_j \leq n$, for $j = 1, \dots, k$.
To make inferences about $\boldsymbol{\theta}$, we consider the model with sampling distribution $\boldsymbol{n}\mid n,\boldsymbol{\theta}\sim\textsf{Mult}( n,\boldsymbol{\theta})$ and the prior distribution $\boldsymbol{\theta}\sim\textsf{Dir}(a_1,\ldots,a_k)$, i.e.,
\begin{equation}\label{eq_dirichlet_1}
	p( \boldsymbol{\theta}) =  \frac{\Gamma \left( \textstyle\sum_{j=1}^{k} a_j \right)}{\textstyle\prod_{j=1}^k \Gamma (a_j)} \prod_{j=1}^k \theta_j^{a_j-1} \,, 
\end{equation}
where $a_1,\ldots,a_k$ are the hyperparameters of the model. Figure \ref{fig_dag_multinomial} shows the model's representation through a directed acyclic graph (DAG).

Using \eqref{eq_multinomial_1} and \eqref{eq_dirichlet_1}, a direct application of Bayes' Theorem states that the posterior distribution of $\boldsymbol{\theta}$ is such that
\begin{align*}
     p(\boldsymbol{\theta}\mid\boldsymbol{n}) \propto p(\boldsymbol{n}\mid \boldsymbol{\theta}) \cdot p(\boldsymbol{\theta}) \propto \prod_{j=1}^k \theta_j^{ n_j } \times \prod_{j=1}^k \theta_j^{a_j-1} =\prod_{j=1}^k  \theta_j^{n_j + a_j-1}\,,
\end{align*}
which corresponds to the kernel of a Dirichlet distribution with parameters $a_1+n_1,\ldots,a_k+n_k$. Therefore, we get that $\boldsymbol{\theta}\mid\boldsymbol{n}\sim\textsf{Dir}(a_1+n_1,\ldots,a_k+n_k)$, i.e., the family of Dirichlet distributions is conjugate to the Multinomial sampling distribution (see Sec. \ref{sec:conjugadas} for more details). 
Finally, we illustrate a typical property of conjugate models. Given that the expected value of the $j$-th component of a random vector with $\textsf{Dirichlet}(c_1,\ldots,c_k)$ distribution is $c_j/c^*$, with $c^* = \sum_{j=1}^k c_j$, the posterior mean of $\theta_j$ is given by 
\begin{align*}
  \textsf{E}(\theta_j \mid \boldsymbol{n})=\frac{a_j+n_j}{\sum_{j=1}^k (a_j+n_j)}=\frac{a_j+n_j}{a^*+n}=\frac{a^*}{a^*+n}\cdot\frac{a_j}{a^*} + \frac{n}{a^*+n}\cdot \frac{n_j}{n}\,,
\end{align*}
where $a^*=\sum_{j=1}^k a_j$ and $n = \sum_{j=1}^k n_j$, and consequently, the posterior mean of $\theta_j$ corresponds to a weighted mean between the prior mean of $\theta_j$ and the sample mean of category $j$, for $j = 1,\ldots,k$.

We present the results of the Multinomial-Dirichlet model fit below. We draw 50000 IID samples of the posterior distribution of $\boldsymbol{\theta}$ to estimate the proportion of votes for Pacto Histórico, Coalición Equipo por Colombia, and Coalición Centro Esperanza. We use $a_1=\ldots=a_k=\frac12$ \citep[this choice of hyperparameters corresponds to Jeffreys' prior;][]{gelman2009bayes}. In Appendix \ref{app_multinomial_dirichlet} we present an algorithm to simulate IID samples from the Dirichlet distribution.

In Table \ref{tab_res_consultas}, we compare our results (posterior mean) with the final report of the Registraduría Nacional del Estado Civil, which is the observed value in Election Day\footnote{https://resultados.registraduria.gov.co/}. We see that for Pacto Histórico and Coalición Centro Esperanza candidates, all credible intervals contain the observed value. On the other hand, for  Coalición Equipo por Colombia, all the intervals, except the one corresponding to candidate David Barguil, do not include the observed value. We strongly believe that this happened because of unexpected political changes prior to Election Day.

\begin{table}[!ht]
\centering
\begin{tabular}{clcccc}
\hline
Consultation & Candidate & Observed & Mean & 2.5\% & 97.5\% \\ 
\hline
 & G. Petro              & 80.50 & 78.18 & 74.08 & 82.02 \\ 
Pacto & F. Márquez       & 14.05 & 13.70 & 10.55 & 17.19 \\
Histórico & C. Romero    & 4.06  & 5.94  & 3.87 & 8.42 \\ 
& A. U. Guariyú          & 0.98  & 1.82  & 0.76 & 3.34 \\ 
& A. Saade               & 0.38  & 0.36  & 0.03 & 1.12 \\ 
\hline
& F. Gutiérrez           & 54.18 & 28.85 & 22.39 & 35.68 \\ 
Coalición &  A. Char     & 17.72 & 24.37 & 18.40 & 30.96 \\ 
Equipo por & E. Peñalosa & 5.80  & 18.77 & 13.42 & 24.84 \\ 
Colombia & D. Barguil    & 15.77 & 15.41 & 10.52 & 21.09 \\ 
& A. Lizarazo            & 6.51  & 12.61 & 8.14 & 17.87 \\ 
\hline
& S. Fajardo             & 33.50 & 37.45 & 29.11 & 46.08 \\ 
Coalición & J. M. Galán  & 22.55 & 23.46 & 16.39 & 31.34 \\
Centro & C. Amaya        & 20.89 & 15.23 & 9.46 & 22.18 \\ 
Esperanza& A. Gaviria    & 15.58 & 12.76 & 7.42 & 19.18 \\
& J. E. Robledo          & 7.46  & 11.11 & 6.13 & 17.30 \\ 
\hline
\end{tabular}
\caption{Observed value, posterior mean, and lower ($2.5\%$) and upper ($97.5\%$) limits of a 95\% confidence interval based on percentiles for each candidate of each political group. Quantities expressed in percentage points.}
\label{tab_res_consultas}
\end{table}

\subsection{Population dynamics: A Poisson regression model}

In this study, we examine the investigation conducted by  \cite{arcese1992stability} on the reproductive activities of $n=52$ female sparrows during the summer. The research was later revisited by\citet[Chap. 10]{hoff2009first}, who applied the Bayesian approach to analyze the data.
We study the number of offspring as a function of age through a Poisson regression model. Although this application is typical of Bio-statistics, it is also interesting from the point of view of Social Sciences because it is strongly related to reproductive patterns and population dynamics.
In this case, we illustrate the Metropolis-Hastings algorithm along with the Hamiltonian Monte Carlo algorithm for obtaining samples from the posterior distribution.

Given that the number of offspring is a count variable, we propose to model this variable as a function of age employing the following model:
\begin{equation}\label{eq_modelo_poisson}
y_i\mid\theta_i\stackrel{\text {iid}}{\sim}\textsf{Poisson}(\theta_i)\,,
\end{equation}
where $y_i$ is the number of offspring of sparrow $i$, for $i=1,\ldots,n$, $\eta_i = \log(\theta_i) = \sum_{j=1}^k\beta_j\, x_{i,j} = \boldsymbol{\beta}^{\textsf{T}}\boldsymbol{x}_i$ is the linear predictor associated with the patterns in the data related to the fixed effects, with $\boldsymbol {\beta}=(\beta_1,\ldots,\beta_k)$ and
$\boldsymbol{x}_i = (x_{i,1},\ldots,x_{i,k})$, and finally, $x_{i,j}$ is the predictor $j$ observed in individual $i$, for $i=1,\ldots,n$ and $j=1,\ldots,k$.
This formulation constitutes a generalized linear model \citep[GLM,][]{mccullagh2018generalized} with a logarithmic link function.

A plot of the number of offspring versus age suggests that number of offspring varies with age according to a concave relationship \citep[p. 172]{hoff2009first}.
For this reason, we specify a linear predictor using a quadratic function of the form
$\eta_i = \beta_1 + \beta_2\,\text{age}_i + \beta_3\,\text{age}^2_i$,
so $k = 3$, $\boldsymbol{\beta}=(\beta_1,\beta_2,\beta_3)$ and $\boldsymbol{x}_i = (x_{i,1},x_{i,2 },x_{i,3})$, with $x_{i,1} = 1$, $x_{i,2} = \text{age}_i$, and $x_{i,3} = \text {age}^2_i$, for $i=1,\ldots,n$.
In addition, we observe that the distribution \eqref{eq_modelo_poisson} may be restrictive since under this formulation, we have that $\textsf{E}(y_i\mid\theta_i) = \textsf{Var}(y_i\mid\theta_i) = \theta_i$. For this reason, we recommended examining the model's goodness of fit through relevant test statistics (see Sec. \ref{sec:ajuste} for more details).
Other popular alternatives to the Poisson distribution are the Negative Binomial distribution (overdispersion: the variation is greater than the expected value) and the Comway-Maxwell-Poisson distribution (underdispersion: the variation is less than the expected value).

\begin{figure}
	\centering
	\includegraphics[scale=.5]{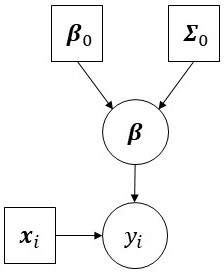}
	\caption{\footnotesize{DAG for the Poisson regression model.}}
	\label{fig_dag_poisson}
\end{figure}

To complete the model specification with sampling distribution \eqref{eq_modelo_poisson}, it is necessary to specify a prior distribution for $\boldsymbol{\beta}$.
Except for the Normal regression model, there are generally no conjugate priors for regression parameters when working with GLMs. However, a standard family of prior distributions that works well in practice is the family of multivariate Normal distributions, so we let $\boldsymbol{\beta}\sim\textsf{N}(\boldsymbol{\beta}_0,\mathbf{\Sigma}_0)$ as a random mechanism to specify the external information about $\boldsymbol{\beta}$.
Consequently, the model parameters are $\beta_1,\ldots,\beta_k$ and the model hyper-parameters are $\boldsymbol{\beta}_0$ and $\mathbf{\Sigma}_0$.
Figure \ref{fig_dag_poisson} shows the representation of the model using a DAG.

In this case, the posterior distribution of $\boldsymbol{\beta}$ is
$$
p(\boldsymbol{\beta}\mid\boldsymbol{y})
\propto \prod_{i=1}^n e^{-\theta_i}\,\theta_i^{y_i} \times \exp\left\{ -\tfrac{1}{2} \boldsymbol{\beta}^{ \textsf{T}}\mathbf{\Sigma}_0^{-1}\boldsymbol{\beta} + \boldsymbol{\beta}^{\textsf{T}}\mathbf{\Sigma}_0^{-1 }\boldsymbol{\beta}_0 \right\}\,,
$$
with $\boldsymbol{y}=(y_1,\ldots,y_n)$ and $\theta_i = \exp{\left(\boldsymbol{\beta}^{\textsf{T}}\boldsymbol{x}_i\right )}$, for $i=1,\ldots,n$, or equivalently in logarithmic scale,
\begin{equation}\label{eq_log_beta_poisson}
\log p(\boldsymbol{\beta}\mid\boldsymbol{y}) = \boldsymbol{\beta}^{\textsf{T}}\sum_{i=1}^n y_i\boldsymbol{x}_i - \sum_{i=1}^n \exp{\left(\boldsymbol{\beta}^{\textsf{T}}\boldsymbol{x}_i\right)}
-\tfrac{1}{2} \boldsymbol{\beta}^{\textsf{T}}\mathbf{\Sigma}_0^{-1}\boldsymbol{\beta} + \boldsymbol{\beta}^{ \textsf{T}}\mathbf{\Sigma}_0^{-1}\boldsymbol{\beta}_0 + \mathrm{C}\,,
\end{equation}
where $\mathrm{C}$ is a constant that does not depend on $\boldsymbol{\beta}$, and consequently the corresponding gradient is 
\begin{equation}\label{eq_log_beta_gradiente_poisson}
    \frac{\partial}{\partial \boldsymbol{\beta}}\log p(\boldsymbol{\beta} \mid \yv)
     = \sum_{i=1}^n \left( y_i - \exp{\left(\boldsymbol{\beta}^{\textsf{T}}\boldsymbol{x}_i\right)} \right)\boldsymbol{x}_i\,.
\end{equation}
We note that $p(\boldsymbol{\beta}\mid\boldsymbol{y})$ does not correspond to any parametric family of standard distributions, which motivates the use of specialized algorithms to explore this posterior distribution through dependent random sequences.
In particular, the Metropolis algorithm and the Hamiltonian algorithm allow us to empirically approximate $p(\boldsymbol{\beta}\mid\boldsymbol{y})$ through a sequence of values $\boldsymbol{\beta}^{(1) },\ldots,\boldsymbol{\beta}^{(B)}$ generated in a Markovian manner (see Sec. \ref{sec:MCMC} for more details). Details about these algorithms are provided in Appendix \ref{app_regresion_poisson}.

In this case, we fit the model assuming a non-informative prior information, by letting $\beta_j\simiid\textsf{N}(0,10)$, for $j=1,2,3$, i.e., $\boldsymbol{\beta}\sim\textsf{N}(\boldsymbol{\beta}_0,\mathbf{\Sigma}_0)$, where $\boldsymbol{\beta}_0 = \boldsymbol{0} _3$ and $\mathbf{\Sigma}_0 = 10\,\mathbf{I}_3$.
We choose as initial value $\boldsymbol{\beta}^{(0)} = \boldsymbol{0}_3$. Then, we run the algorithms using $10000$ iterations after a warm-up period of $1000$ iterations.
On the one hand, in order to implement the Metropolis-Hastings algorithm, we use $\mathbf{\Delta}_0 = \mathrm{c}\,(\mathbf{X}^{\textsf{T}}\mathbf{X})^{- 1}$, with $\mathrm{c} = 0.7$ and $\mathbf{X} = [\boldsymbol{x}_1,\dots,\boldsymbol{x}_n]^{\textsf{T}}$. On the other hand, in order to implement the Hamiltonian Monte Carlo algorithm, we use $L = 100$, $\epsilon = 0.01$, and $\mathbf{M}=\mathbf{I}_3$. These adjustments lead to favorable acceptance rates of $38\%$ and $66\%$, respectively (see \citealp[Chap. 12]{gelman2013bayesian} for more details about the selection of tunning parameters).

\begin{figure}[!ht]
	\centering
	\setlength{\tabcolsep}{0pt}
	\begin{tabular}{ccc}
    & \hspace{0.35cm} Metropolis & \hspace{0.15cm} Hamiltonian \\
	\begin{sideways} \hspace{2.2cm} $\beta_1$ \end{sideways}   &
	\includegraphics[scale=.5]{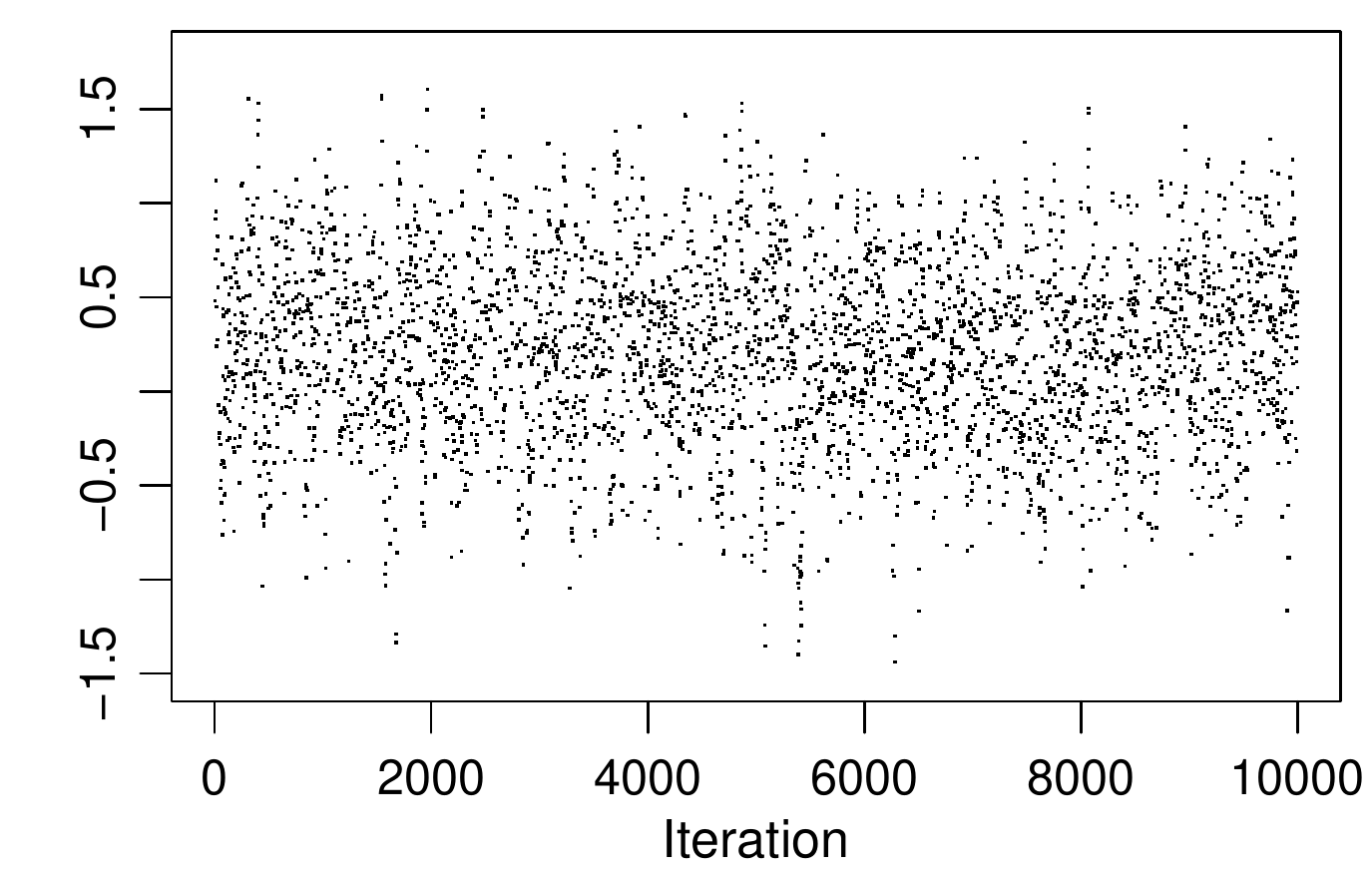} &
	\includegraphics[scale=.5]{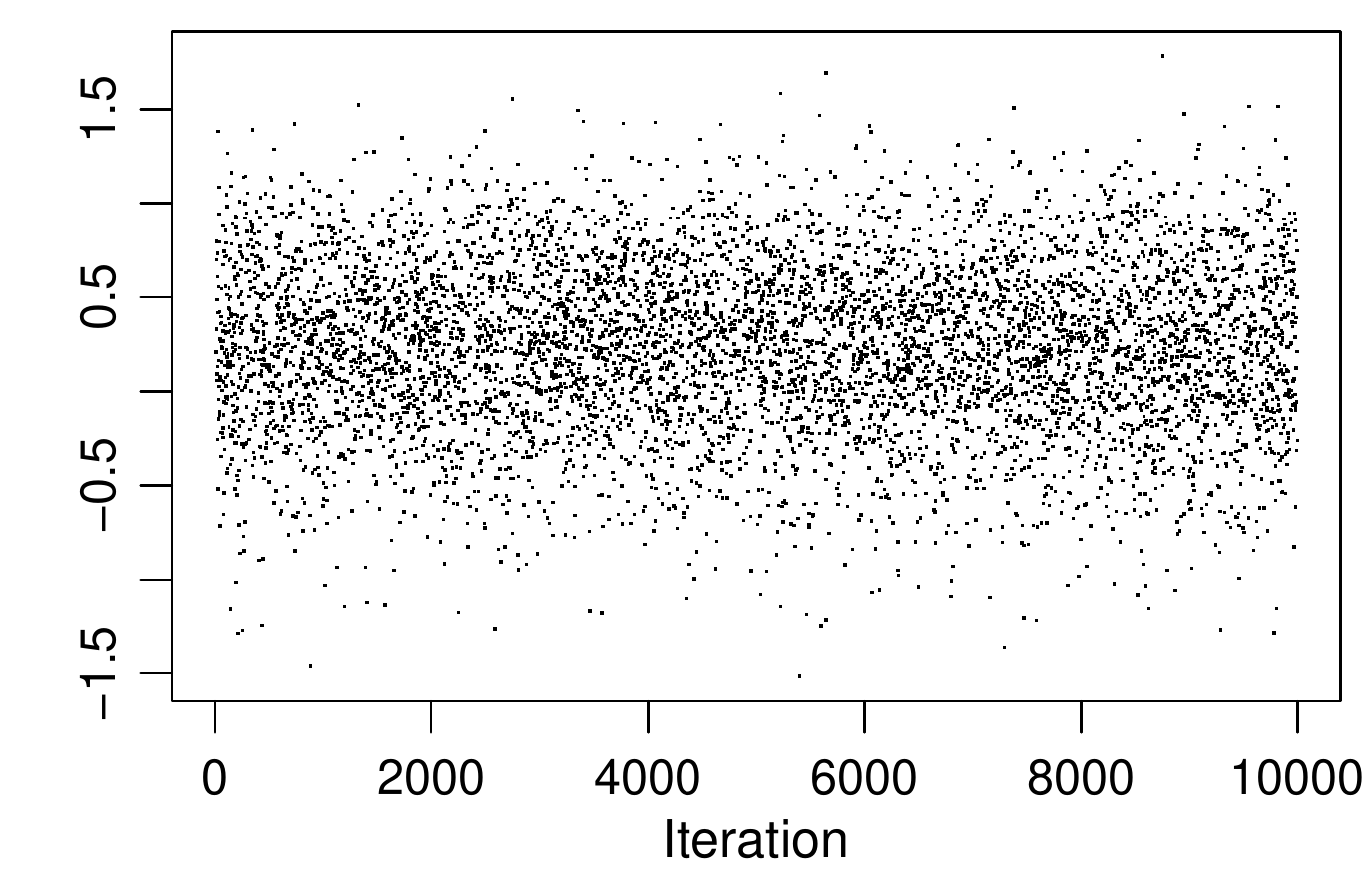}   \\
	\begin{sideways} \hspace{2.2cm} $\beta_2$ \end{sideways}   &
	\includegraphics[scale=.5]{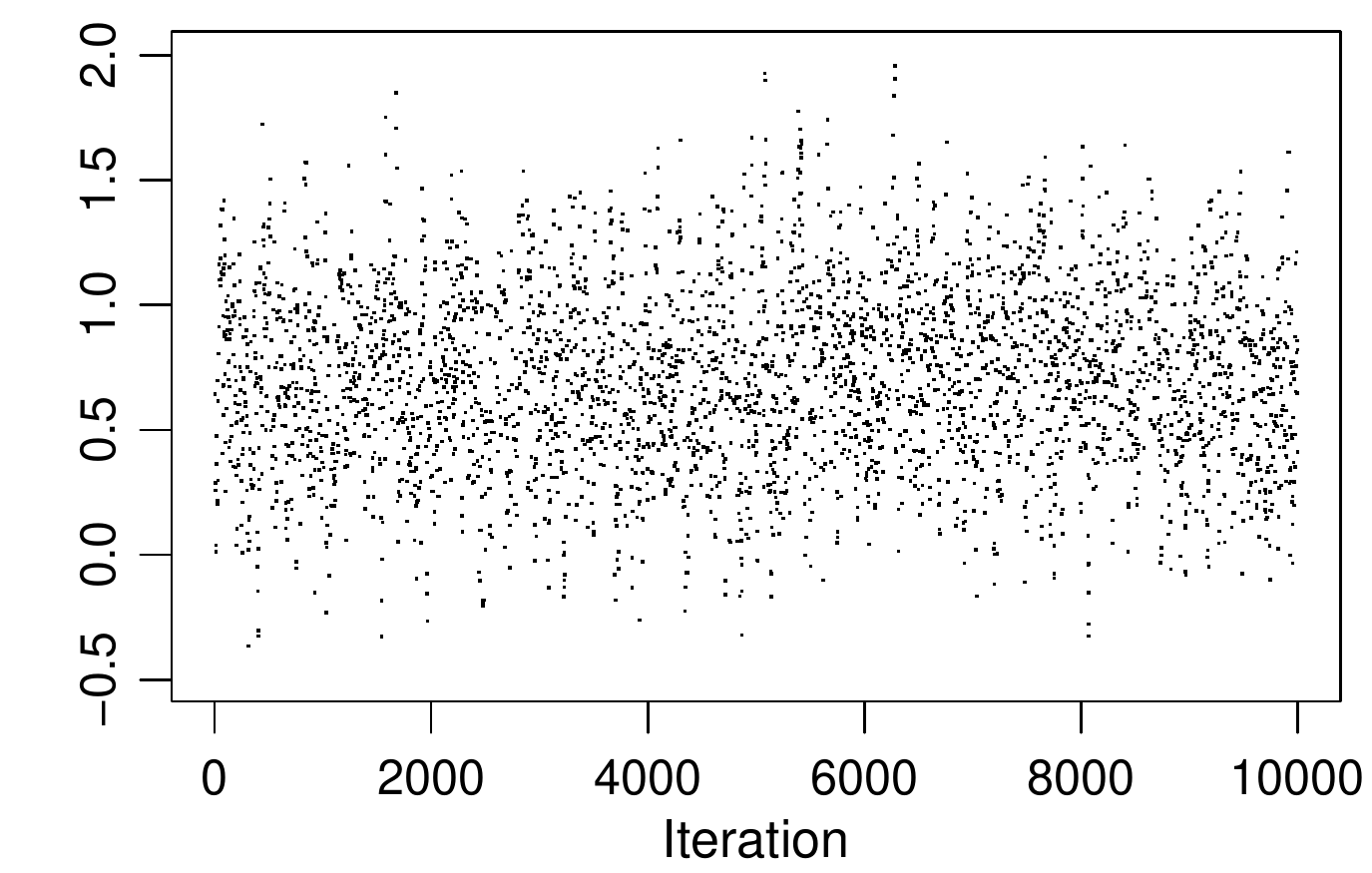} &
	\includegraphics[scale=.5]{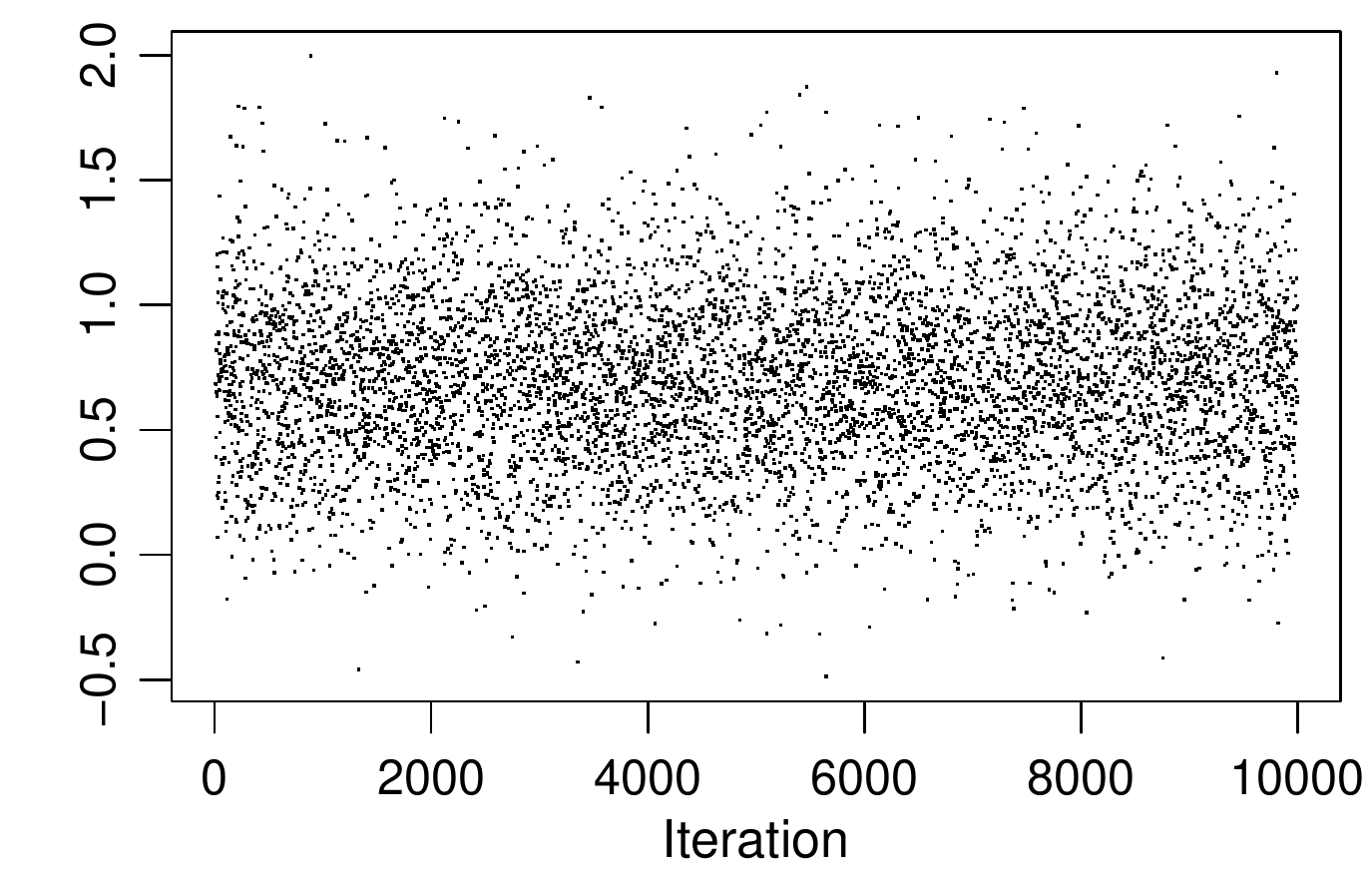}   \\
	\begin{sideways} \hspace{2.2cm} $\beta_3$ \end{sideways}   &
	\includegraphics[scale=.5]{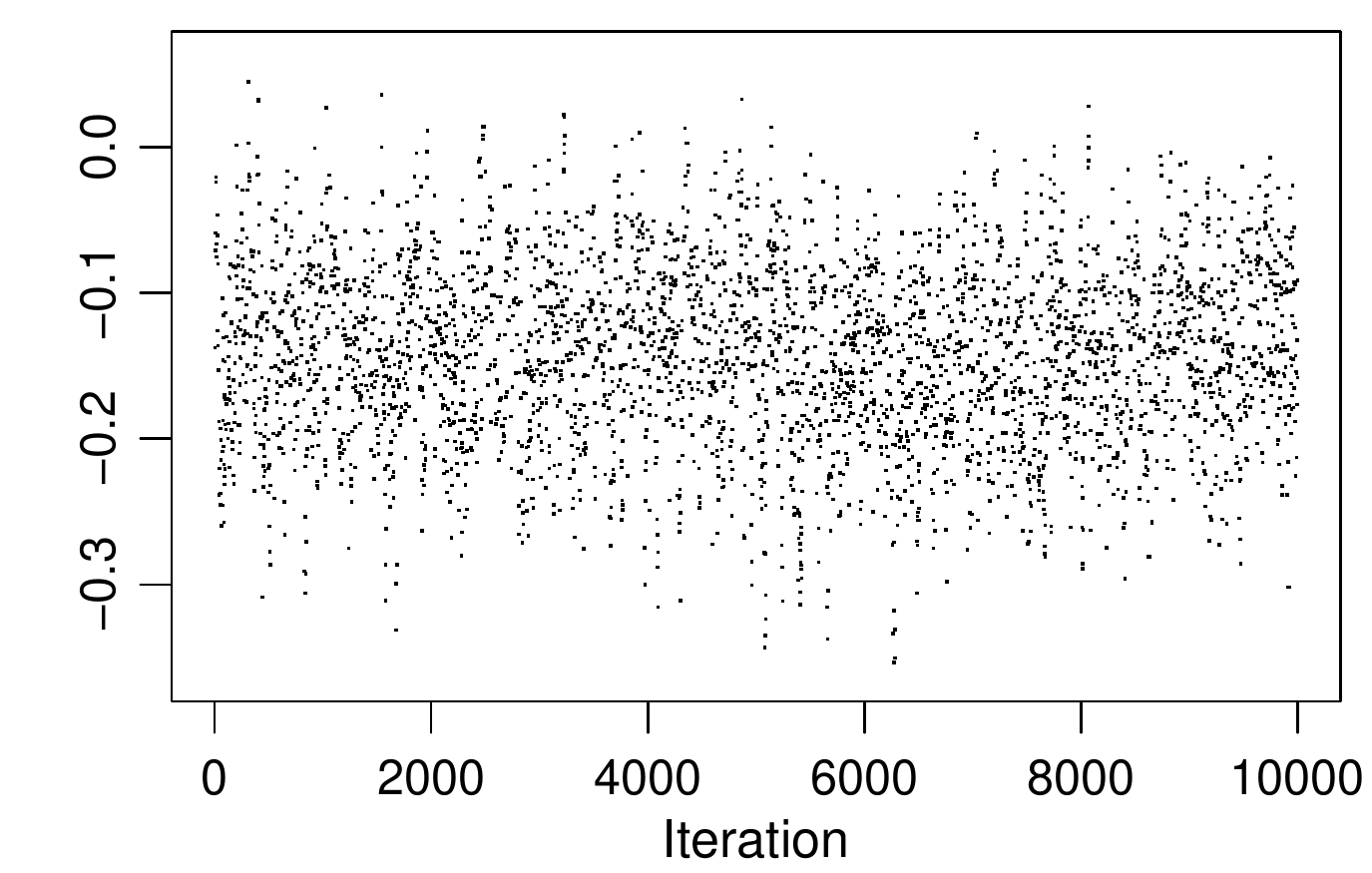} &
	\includegraphics[scale=.5]{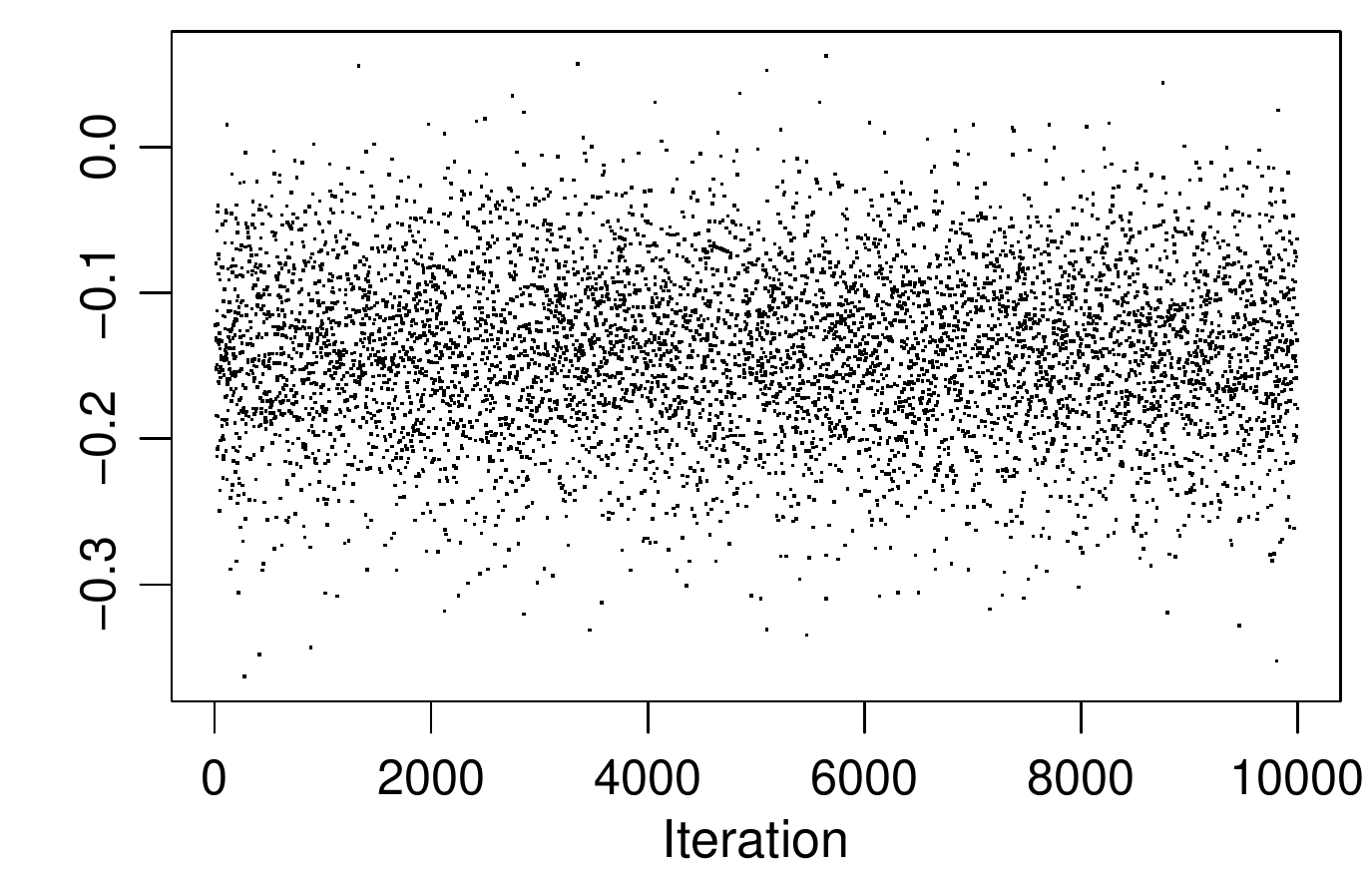}   \\
	\end{tabular}
	\caption{\footnotesize{Markov chains associated with the posterior distribution $p(\boldsymbol{\beta}\mid\boldsymbol{y})$ of the Poisson regression model.}}
	\label{fig_cadenas_poisson}
\end{figure}

Figure \ref{fig_cadenas_poisson} shows the Markov chains associated with $p(\boldsymbol{\beta}\mid\boldsymbol{y})$. We observe no evidence of a lack of convergence. Furthermore, we notice that the Hamiltonian algorithm produces chains with better mixing properties than the Metropolis-Hastings algorithm (this is expected given that the Hamiltonian's convergence rate is higher than Metropolis's). Finally, both the effective sample sizes and the Monte Carlo errors presented in Table \ref{tab_convergencia_poisson} confirm that these chains are appropriate to make inferences about the parameters of interest (again, it is evident that the Hamiltonian algorithm is more efficient in exploring the posterior distribution).

% latex table generated in R 4.2.1 by xtable 1.8-4 package
% Fri Aug 26 11:37:11 2022
\begin{table}[!ht]
\centering
\begin{tabular}{ccccc}
  \hline
  \multirow{2}{*}{Parameter} & \multicolumn{2}{c}{Effective size} & \multicolumn{2}{c}{MC error} \\
  \cline{2-5}
  & Metropolis & Hamiltonian & Metropolis & Hamiltonian \\	
  \hline
  $\beta_1$ & 802.2 & 6429.1 & 0.016 & 0.005 \\ 
  $\beta_2$ & 729.1 & 6407.7 & 0.013 & 0.004 \\ 
  $\beta_3$ & 665.7 & 6233.2 & 0.002 & 0.001 \\ 
   \hline
\end{tabular}
\caption{\footnotesize{Effective sample sizes and Monte Carlo errors corresponding to the Markov chains associated with the posterior distribution $p(\boldsymbol{\beta}\mid\boldsymbol{y})$ of the Poisson regression model.}}
	\label{tab_convergencia_poisson}
\end{table}

\begin{figure}[!ht]
    \centering
    \subfigure[$p(\beta_1\mid\yv)$]  {\includegraphics[scale=.6]{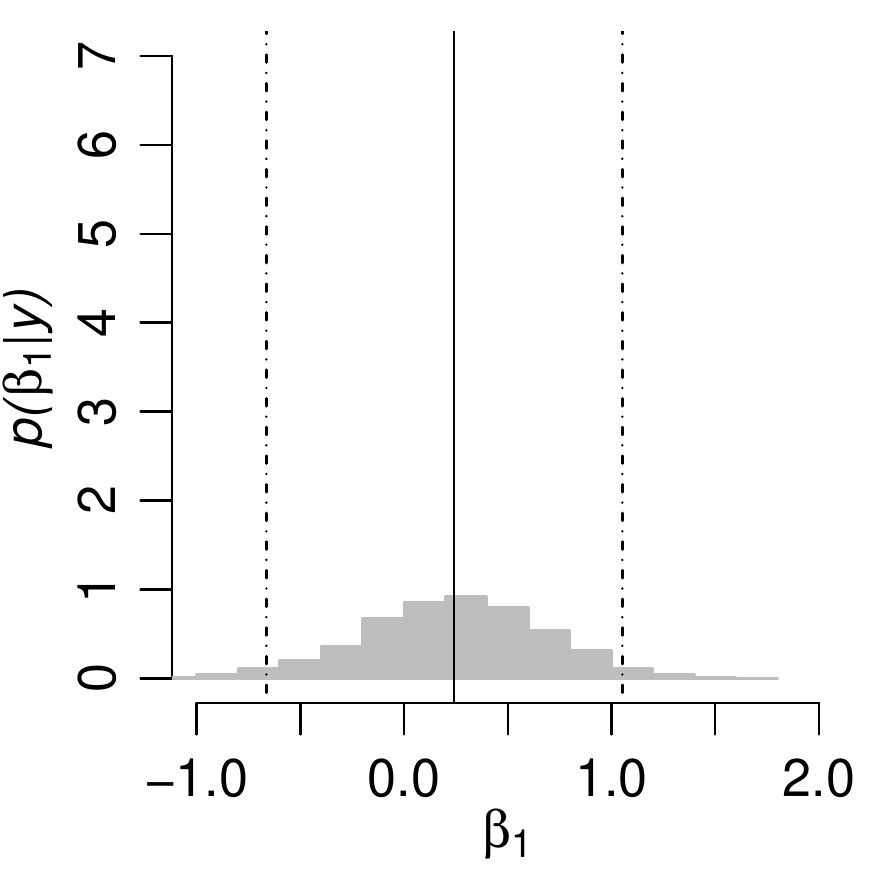}}
    \subfigure[$p(\beta_2\mid\yv)$]  {\includegraphics[scale=.6]{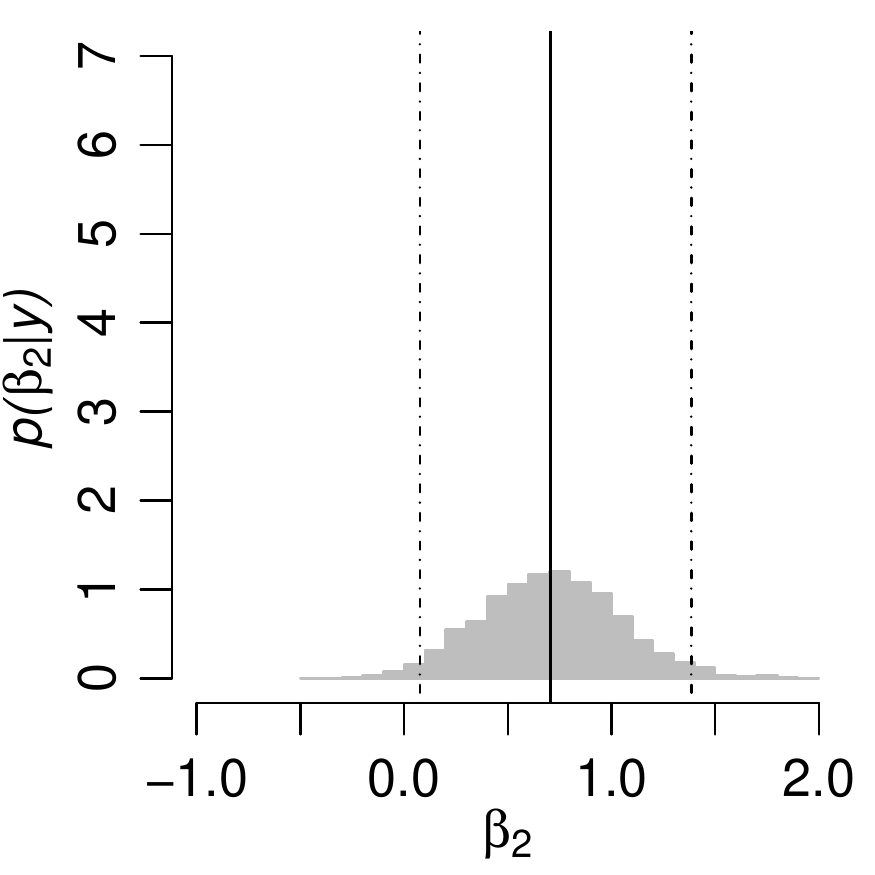}}
    \subfigure[$p(\beta_3\mid\yv)$]  {\includegraphics[scale=.6]{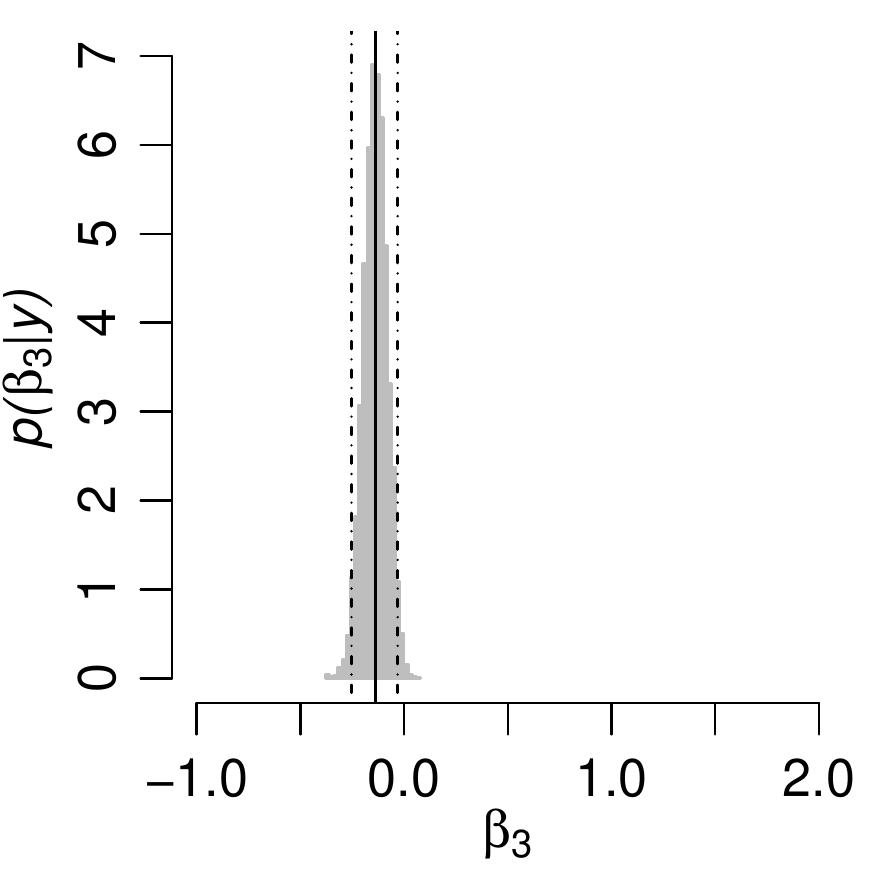}}\\
    \subfigure[Number of pups]      {\includegraphics[scale=.6]{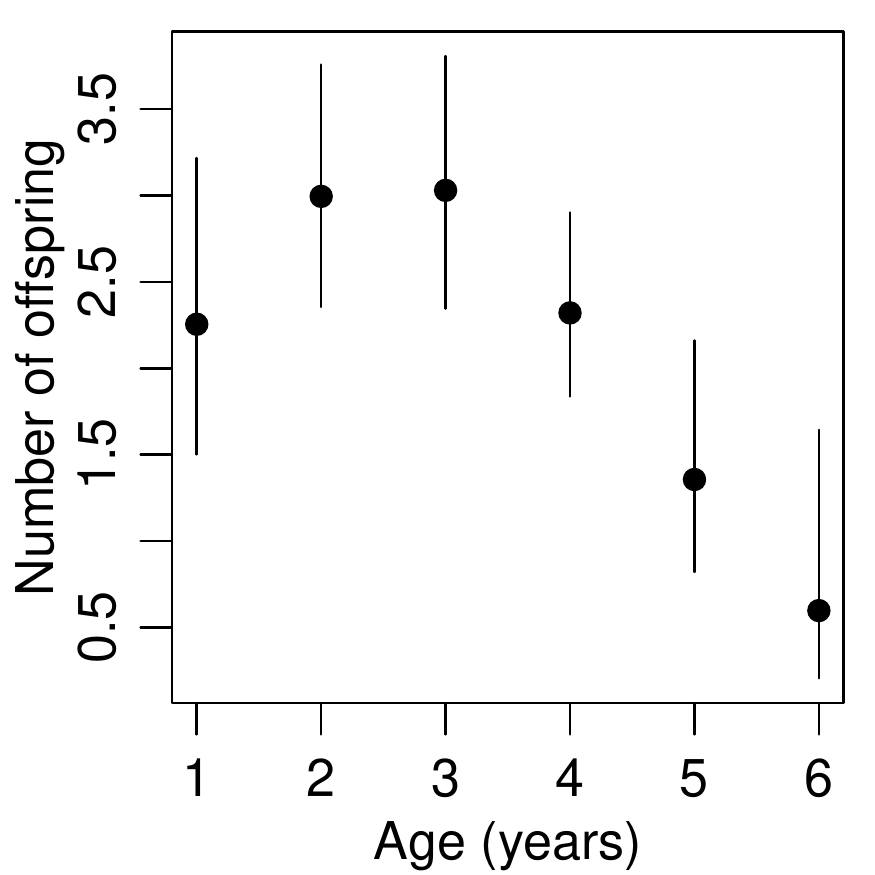}}
    \subfigure[Mean]                {\includegraphics[scale=.6]{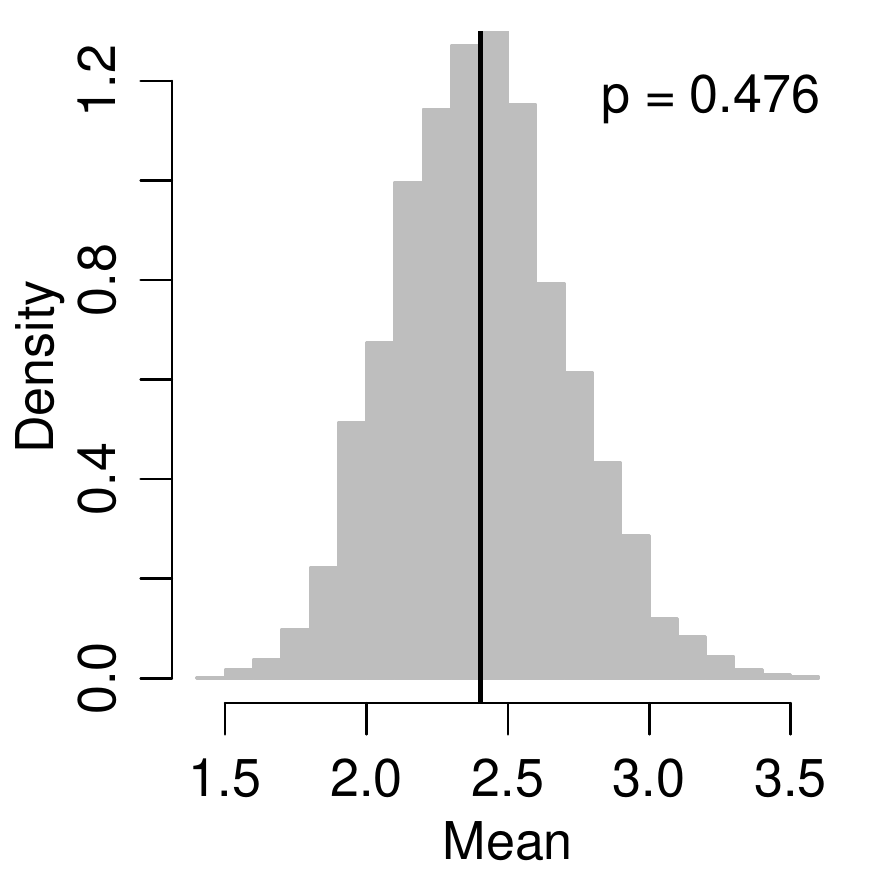}}
    \subfigure[Variance]             {\includegraphics[scale=.6]{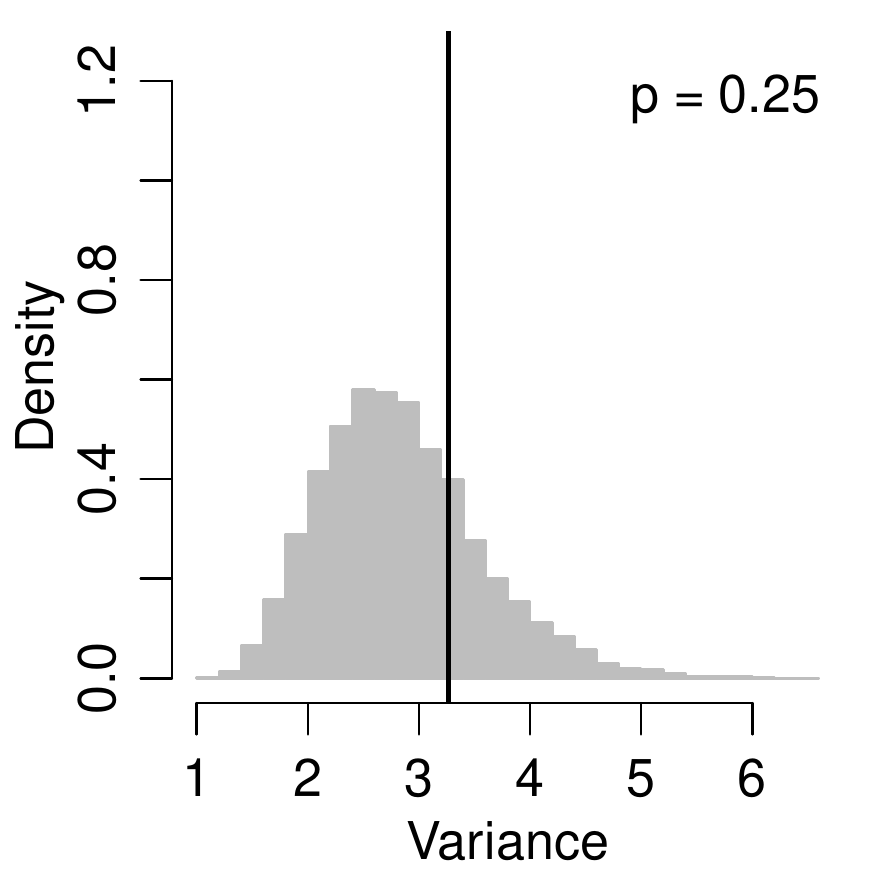}}
    \caption{\footnotesize{Panels (a)-(c): posterior distribution of $\beta_1$, $\beta_2$, and $\beta_3$, along with the mean posterior (solid line) and the limits of a 95\%  credibility interval based on percentiles (dotted lines). Panel (d): posterior mean and limits of a 95\% credibility interval based on percentiles, for age in $\{1;\ldots;6\}$. Panels (e)-(f): posterior predictive distribution of the mean and variance (test statistics), along with the observed value (solid line) and the corresponding posterior predictive $p$ value.}}
    \label{fig_post_poisson}
\end{figure}

Figures (a), (b), and (c) in Figure \ref{fig_post_poisson} display the posterior distribution of the regression coefficients, accompanied by the respective point estimate and a 95\% credible interval based on percentiles.
Our findings indicate that age and age-squared effects are significant (credible intervals do not contain 0).
Furthermore, the signs of the point estimates of $\beta_2$ (positive) and $\beta_3$ (negative) confirm that the number of offspring varies with age through a concave relationship.
This behavior is clear in panel (d) of Figure \ref{fig_post_poisson}, where it is evident that the reproductive pattern of this species has a moderate period of ascent (years 1 and 2), then reaches a peak (year 3), and then, it has a prolonged period of decline (years 3 to 6).

Finally, the model's goodness of fit is evaluated by means of the posterior predictive distribution of a perdifined set of test statistics (see Sec. \ref{sec:ajuste} for more details). In this case, the mean and variance are chosen as test statistics since they characterize essential aspects of the data (trend and dispersion) that might be overshadowed due to the mean-variance restriction of the Poisson model. Panels (e) and (f) of Figure \ref{fig_post_poisson} suggest that the model fits the data  well because the observed values of the data are typical values of the posterior predictive distribution of the corresponding test statistics (i.e., posterior predictive $p$ values are not close to either 0 or 1).

\subsection{Standardized educational testing: Hierarchical linear regression model}

In this study, we employed three multiple linear regression models to examine the math score outcomes of the Saber 11 Test during the first semester of 2020 in Colombia.
The Instituto Colombiano para la Evaluación de la Educación (ICFES) applies this standardized test periodically to measure the skills of students who finish secondary school. We aim to make inferences about the Colombian student population at the national and departmental levels about their performance in mathematics. We examined the score in mathematics because it is a variable that social researchers usually relate to other important educational factors \citep[e.g.,][]{anis2016relations,vzivkovic2023math}. This dataset is publicly available\footnote{\url{https://www2.icfes.gov.co/data-icfes}}.

Based on the exam design, production, application, and scoring guide of the Saber 11 exam, the mathematics test is graded on a scale ranging from 0 to 100, with whole numbers only. The average score is set at 50 points, with a standard deviation of 10 points. In our analysis, we treated the score as the response variable, while considering the student's sex and employment status as covariates. 
Prior to model fitting, a pre-processing step was conducted, which involved eliminating all records with missing data. Furthermore, the variables ``sex'' and ``employment status'' were encoded (sex: 1 if male, 0 if female; employment status: 1 if worked 0 hours during the last week, 0 otherwise). 
The resulting dataset, formed through these adjustments, comprised a total of 14,015 records.
Bayesian imputation methods are available \citep[see for example Ch. 7][]{hoff2009first}. However, since the percentage of records lost by direct deletion is very small, adding this level of additional complexity in any of the models is not necessary.

\begin{figure}[!ht]
    \centering
    \includegraphics[scale=0.65]{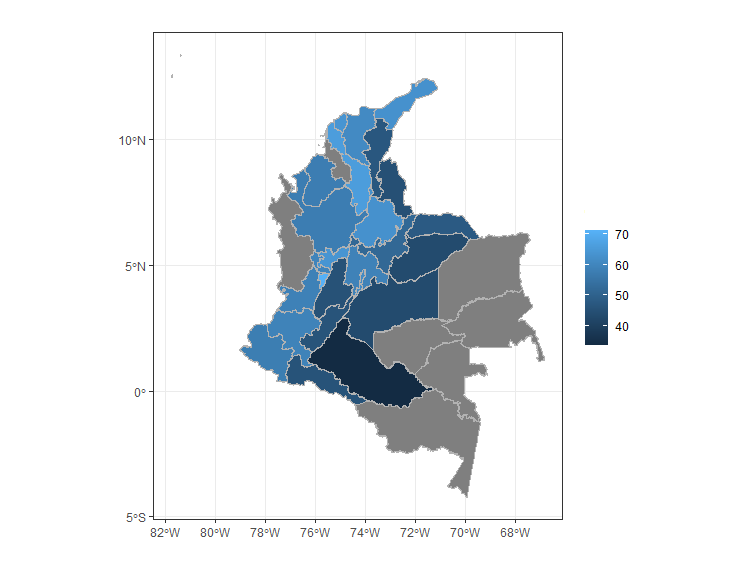}
    \caption{{\footnotesize Average math scores by departments. Departments in gray do not have information regarding math scores.}}\label{mapa}
\end{figure}

Figure \ref{mapa} shows the math sample means of each department. We observe that the averages oscillate between 35 and 70 approximately. In addition, we do not have any information for eight departments (including the archipelago of San Andrés, Providencia, and Santa Catalina).
We also appreciate that the department with the highest average is Quindio, while the lowest is Caquetá. Finally, those departments located in the Orinoquía and Amazonía Regions of the country exhibited the lowest scores nationwide.

\begin{figure}
	\centering
	\includegraphics[scale=.54]{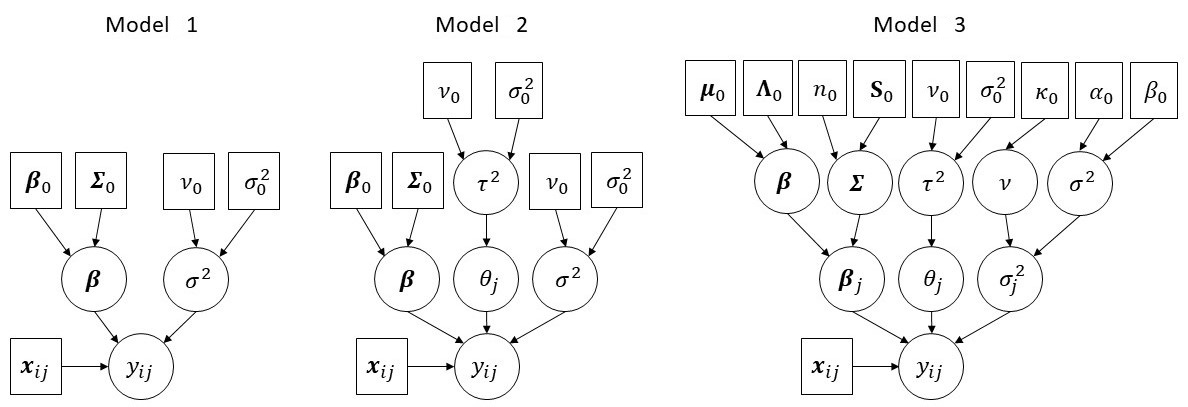}
	\caption{\footnotesize{DAGs for the multiple regression models.}}
	\label{fig_dag_normal}
\end{figure}

Let $y_{i,j}$ and $\boldsymbol{x}_{i,j} = (x_{i,j,1},\ldots,x_{i,j,p})$ be the response variable and the vector of covariates corresponding to individual $i$ in group $j$, respectively, for $i=1,\ldots,n_j$ and $j = 1,\ldots,m$.
In this case, $y_{i,j}$ corresponds to the mathematics score of student $i$ in department $j$, where $n_j$ is the number of students in department $j$, and $m$ is the number of departments.
In addition, $p=3$ covariates are considered, namely $x_{i,j,1}$, constant variable equal to 1 associated with the intercept of the linear predictor, $x_{i,j,2}$, dummy variable associated with the sex of student $i$ in department $j$, and $x_{i,j,3}$, dummy variable associated with the employment condition of student $i$ in department $j$. Three multiple regression models with different characteristics are proposed below to analyze the data. 
Figure \ref{fig_dag_normal} shows the representation of the models using DAGs.
These models can be easily extended to consider spatial information.
The Bayesian paradigm is particularly useful in such a case \citep[e.g.,][]{banerjee2014hierarchical}.

\textbf{Model 1: Multiple linear regression}

\begin{itemize}
\item Sampling distribution:
$$
y_{i,j}\mid\bev,\sig^2,\xv_{i,j} \simind \normal(\xv_{i,j}^{\trans}\bev,\sig^2)\, ,\qquad
 i=1,\ldots,n_j\,,\qquad j = 1,\ldots,m\,,
$$
where $\bev=(\beta_1,\ldots,\beta_p)$ is the vector of regression coefficients and $\sigma^2$ is the variance of the response variable. The sampling distribution is equivalent to
$$
\yv\mid\bev,\sig^2 \sim \normal_n(\Xm\bev,\sig^2\I_n)\,,
$$
where $\yv = (\yv_1,\ldots,\yv_m)$, with $\yv_j = (y_{1,j},\ldots,y_{n_j,j})$, and $\Xm = [\Xm_1 ^{\trans},\ldots,\Xm_m^{\trans}]^{\trans}$, with $\Xm_j = [\xv_1,\ldots,\xv_{n_j}]^{\trans}$, and $\I_n$ is the identity matrix $n\times n$.
\item Prior distribution:
$$
\bev \sim \normal (\bev_0, \SIG_0)\,,\qquad \sig^2 \sim \GI \left(\tfrac{\nu_0}{2},\tfrac{\nu_0\sigma^2_0}{ 2}\right)\,.
$$
\item Hyperparameters: $\bev_0$, $\SIG_0$, $\nu_0$, $\sigma^2_0$.
\end{itemize}

\textbf{Model 2: Multiple linear regression with random effects}

\begin{itemize}
\item Sampling distribution:
$$
y_{i,j}\mid\bev,\theta_{j},\sig^2 \simind \normal(\xv_{i,j}^{\trans}\bev + \theta_{j},\sig^ 2)\,,\qquad
i=1,\ldots,n_j\,,\qquad j = 1,\ldots,m\,,
$$
where $\theta_j$ is the random effect corresponding to group $j$. Here, the random effects $\theta_1,\ldots,\theta_m$ represent the latent (unobserved) characteristics of the groups associated with the response variable. The sampling distribution is equivalent to
$$
\yv\mid\bev,\tev,\sig^2 \sim \normal_n(\Xm\bev + \bs{\vartheta},\sig^2\I_{n})\,,
$$
where $\tev = (\theta_1,\ldots,\theta_m)$ and $\bs{\vartheta} = (\theta_1\bs{1}_{n_1},\ldots,\theta_m\bs{1}_{ n_m})$, with $\bs{1}_{n_j}$ the vector of ones of size $n_j$.
\item Prior distribution:
$$
\theta_{j}\mid\tau^2 \simiid \normal(0,\tau^2)\,,\qquad
\tau^2 \sim \GI\left(\tfrac{\eta_0}{2},\tfrac{\eta_0\tau^2_0}{2}\right)\,,\qquad
\bev \sim \normal (\bev_0, \SIG_0)\,,\qquad
\sig^2 \sim \GI\left(\tfrac{\nu_0}{2},\tfrac{\nu_0\sigma^2_0}{2}\right)\,.
$$
\item Hyperparameters: $\eta_0$, $\tau^2_0$, $\bev_0$, $\SIG_0$, $\nu_0$, $\sigma^2_0$.
\end{itemize}

\textbf{Model 3: Multilevel multiple linear regression with random effects}

\begin{itemize}
\item Sampling distribution:
$$
y_{i,j}\mid\bev_j,\theta_{j},\sig_j^2 \simind \normal(\xv_{i,j}^{\trans}\bev_j + \theta_{j},\sig_j^ 2)\,,\qquad
i=1,\ldots,n_j\,,\qquad j = 1,\ldots,m\,,
$$
where $\bev_j = (\beta_{1,j},\ldots,\beta_{p,j})$ and $\sigma^2_j$ are the vector of regression coefficients and the variance of the response variable corresponding to the group $j$, respectively.
The sampling distribution is equivalent to
$$
\yv_j\mid\bev_j,\tev,\sig_j^2 \simind \normal_{n_j}(\Xm_j\bev_j + \theta_j\bs{1}_{n_j},\sig_j^2\I_{n_j})\,,\qquad j = 1,\ldots,m\,.
$$
\item Prior distribution:
\begin{align*}
\theta_{j}\mid\tau^2 &\simiid \normal(0,\tau^2)\,, &
\bev_j\mid\bev, \SIG &\simiid \normal_n (\bev, \SIG)\,, &
\sig_j^2 \mid \nu,\sig^2 &\simiid \GI \left( \tfrac{\nu}{2}, \tfrac{\nu\sig^2}{2} \right)\,, \\
\tau^2 &\sim \GI\left(\tfrac{\eta_0}{2},\tfrac{\eta_0\tau^2_0}{2}\right)\,, &
\bev &\sim \normal_n(\bs{\mu}_0,\mathbf{\Lambda}_0)\,, &
\nu &\sim e^{-\kappa_0\nu}\,, \\
       &&
\SIG &\sim \textsf{WI}(n_0, \Sm^{-1}_0)\,, &
\sig^2 &\sim \textsf{G}(\alpha_0, \beta_0)\,.
\end{align*}
\item Hyperparameters: $\eta_0$, $\tau^2_0$, $\muv_0$, $\LAM_0$, $n_0$, $\Sm_0$, $\kappa_0$, $\alpha_0$, $\beta_0$.
\end{itemize}

We fit the models using a Gibbs sampler (see Sec. \ref{sec:MCMC} for more details) with $55000$ iterations. The first $5000$ iterations of the algorithm constitute the warm-up period, so they are not considered to carry out the posterior computations. Details of the Gibbs sampler for each model are given in Appendix \ref{app_regresion_lineal_multiple}. Furthermore, we implement the models using the following hyperparameters based on a unit information prior distribution \citep{kass1996selection} as follows:
\begin{itemize}
    \item Model 1: $\bev_0 = \hat\bev_{\textsf{ols}}$, $\SIG_0 = n\,\hat\sig_{\textsf{ols}}^2(\Xm^{\trans} \Xm)^{-1}$, $\nu_0 = 1$, $\sigma^2_0 = \hat\sig_{\textsf{ols}}^2$, where $\hat\bev_{\textsf{ols} }$ and $\hat\sig_{\textsf{ols}}^2$ are the ordinary least squares estimators of $\bev$ and $\sigma^2$, respectively, i.e., $\hat\bev_{\textsf{ols}} = (\Xm^{\trans}\Xm)^{-1}\Xm^{\trans}\yv$ and $\hat\sig_{\textsf{ols}}^2 = \tfrac{ 1}{n-p} (\yv - \Xm\hat\bev_{\textsf{ols}})^{\trans} (\yv - \Xm\hat\bev_{\textsf{ols}})\,.$
    \item Model 2: $\bev_0 = \hat\bev_{\textsf{ols}}$, $\SIG_0 = n\,\hat\sig_{\textsf{ols}}^2(\Xm^{\trans} \Xm)^{-1}$, $\nu_0 = \eta_0 = 1$, $\sigma^2_0 = \tau^2_0 = \hat\sig_{\textsf{ols}}^2$.
    \item Model 3: $\muv_0 = \hat\bev_{\textsf{ols}}$, $\LAM_0 = \Sm_0 = n\,\hat\sig_{\textsf{ols}}^2(\Xm^{ \trans}\Xm)^{-1}$, $n_0 = 5$, $\eta_0 = \kappa_0 = \alpha_0 = 1$, $\tau^2_0 = \beta_0 = \hat\sig_{\textsf{ols}}^2$.
\end{itemize}
An exhaustive convergence analysis (we do not present it here) indicates no signs of  lack of convergence in any case.

Table \ref{tab_post_normal} shows the estimate and 95\% credible intervals based on percentiles for the components of $\bev$ and $\sigma$ for each model ($\bev$ and $\sigma ^2$ are part of the first hierarchy in Models 1 and 2, while the second in Model 3; see Figure \ref{fig_dag_normal}).
On the one hand, the estimates of $\beta_1$ agree with the design of the test. The biggest difference is only $(51.28-50)/50=2.56\%$ regarding the test design (50 points), which in practical terms does not correspond to a substantial difference.
On the other hand, the estimates of $\sigma$ indicate that the variability of the average scores turns out to be significantly higher than the test design (10 points). The smallest difference is $(11.36-10)/10 = 13.6\%$, and the limits of all the credible intervals are greater than 10, which indicates a significantly higher heterogeneity in math scores than initially anticipated.
Finally, the estimates associated with $\beta_2$ and $\beta_3$ indicate that there is a significant effect of gender and employment status on the average math score since the limits of the corresponding credible intervals are greater than 0 (except for that of $\beta_2$ in Model 3). Specifically, being a man working 0 hours a week are characteristics corresponding to a significant increase in the average math score of 3.13 and 8.99 points according to Model 1, 3.05 and 8.17 points according to Model 2, and 2.32 and 6.97 points according to Model 3, respectively.

\begin{table}[!ht]
\centering
\begin{tabular}{cccccccccc}
  \hline
  \multirow{2}{*}{Parameter} & \multicolumn{3}{c}{Model 1} & \multicolumn{3}{c}{Model 2} & \multicolumn{3}{c}{Model 3}\\
  \cline{2-10}
  & Mean & 2.5\% & 97.5\% & Mean & 2.5\% & 97.5\% & Mean & 2.5\% & 97.5\% \\	
  \hline
  $\beta_1$ & 51.28 & 50.83 & 51.74 & 48.79 & 45.51 & 52.02 & 49.35 & 44.93 & 53.7  \\ 
  $\beta_2$ & 3.13  & 2.71  & 3.56  & 3.05  & 2.64  & 3.47  & 2.32  & -0.38 & 5.02  \\
  $\beta_3$ & 8.99  & 8.53  & 9.45  & 8.17  & 7.72  & 8.65  & 6.97  & 3.63  & 10.25 \\ 
  $\sigma$  & 12.76 & 12.61 & 12.91 & 12.43 & 12.29 & 12.58 & 11.36 & 10.07 & 12.66 \\
  \hline
\end{tabular}
\caption{\footnotesize{Posterior mean and lower (2.5\%) and upper (97.5\%) limits of a credibility interval based on 95\% percentiles for the $\bev$ and $\sigma$ components in each model.}}
\label{tab_post_normal}
\end{table}

Table \ref{tab_dic_normal} presents the DIC for each model (see Sec. \ref{sec:comparar} for more details).
The DIC evaluates the model's predictive quality penalizing for the effective number of parameters. The results show that Model 3 has the best predictive capabilities.
Unlike Models 1 and 2, Model 3 is a multilevel model with regression coefficients and specific variance components, which allows internal characterization of each department's dynamics and direct department comparisons.
For this reason, we use Model 3 to analyze behavior and differences between departments.

\begin{table}[!htb]
\centering
\begin{tabular}{lccc}
  \hline
    & Model 1 & Model 2 & Model 3 \\ 
  \hline
   DIC & 111139.6 & 110435.8 & 109982.9 \\ 
   \hline
\end{tabular}
\caption{Deviance information criteria for each model.}
\label{tab_dic_normal}
\end{table}

\begin{figure}[!t]
    \centering
    \subfigure[$\beta_{1,j}$: intercept]         {\includegraphics[scale=.46]{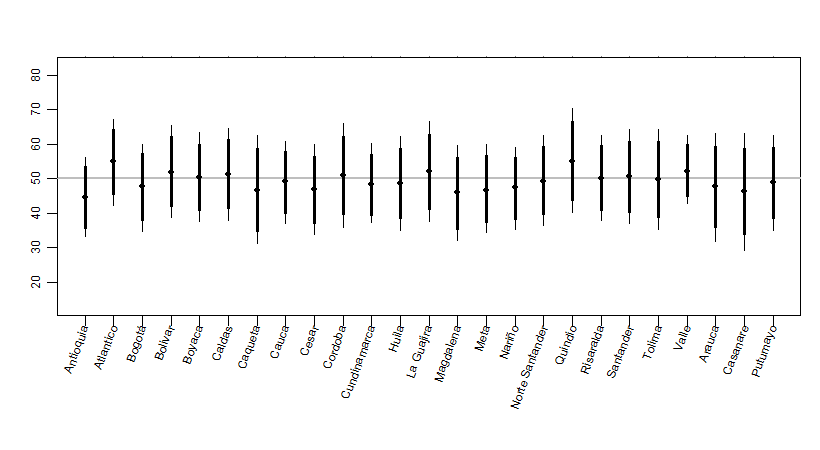}}
    \subfigure[$\beta_{2,j}$: sex]               {\includegraphics[scale=.46]{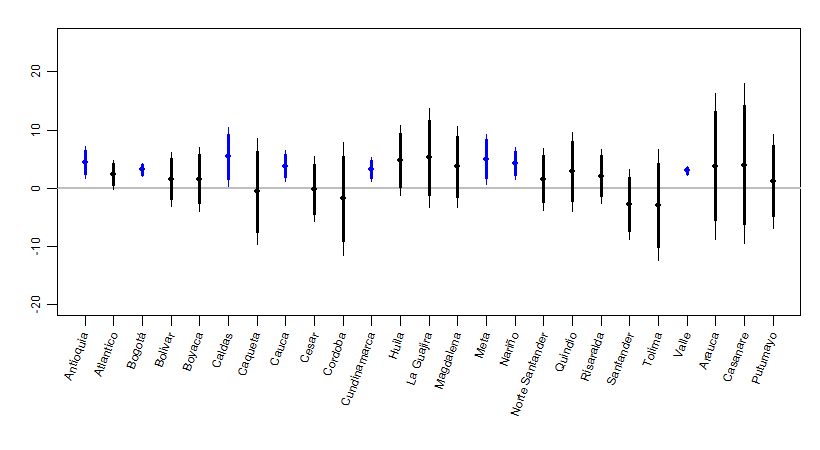}}
    \subfigure[$\beta_{3,j}$: employment status]  {\includegraphics[scale=.46]{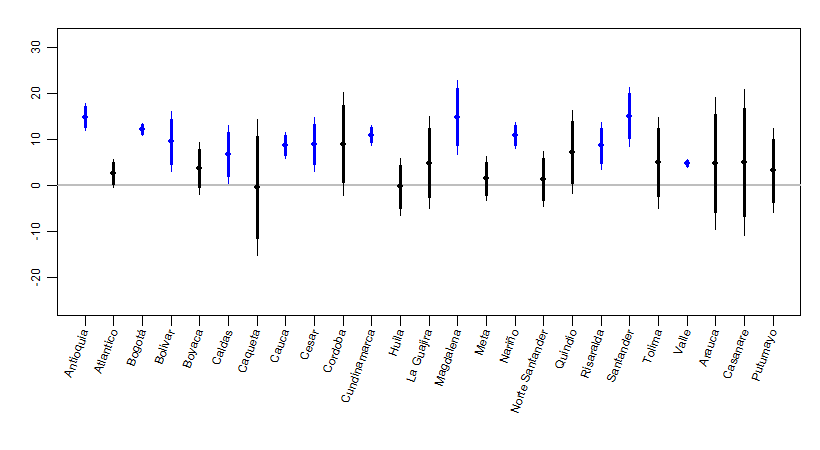}}
    \caption{\footnotesize{Posterior mean and credible intervals based on percentiles using 95\%  (thick lines) and 99\% (thin line) confidence, for each regression coefficient $\boldsymbol{\beta}_{k,j}$, with $k \in \{1,2,3\}$ and $j \in \{1,\cdots, 25\}$. Intervals in blue do not contain the reference value (50 for $\beta_1$ and 0 for $\beta_2$ and $\beta_3$).}}
    \label{fig_inferencia_normal}
\end{figure}

Figure \ref{fig_inferencia_normal} shows the posterior means and credible intervals based on percentiles using 95\% and 99\% confidence, for each regression coefficient $\boldsymbol{\beta}_{k,j}$, with $k \in \{1,2,3\}$ and $j \in \{1,\cdots, 25\}$. These plots allow us to identify trends and significant differences from the reference values (50 for $\beta_1$ and 0 for $\beta_2$ and $\beta_3$) across departments.
Intervals in blue do not contain the reference value, indicating significant differences concerning the corresponding reference value.
Panel (a) of Figure \ref{fig_inferencia_normal} indicates that all the departments behave very similarly concerning the intercept, given that all the posterior means are close to 50 and all the credibility intervals contain this value. This phenomenon confirms the suitability of the test design in terms of central tendency.
On the other hand, panel (b) of Figure \ref{fig_inferencia_normal} indicates significant differences regarding sex respect to the reference value in eight departments.
This empirical evidence is unfortunate in terms of equity because the sex of the individual is not expected to have a significant association on the individual's performance on the test.
Indeed, this is the case in Antioquia, Bogotá, Caldas, Cauca, Cundinamarca, Meta, Nariño, and Valle del Cauca, where there exists a significant increase in the math score in favor of men.
Finally, panel (c) of Figure \ref{fig_inferencia_normal}, once again indicates significant differences regarding employment status respect to the reference value, but this time in 12 departments.

Interestingly, employment status is significant in those departments where sex is also significant (except in Meta), in all cases, in favor of those individuals who did not work the week before taking the test.
Other departments that turned out to have a significant association concerning employment status are Magdalena, Santander, and Risaralda.
We observe that the most developed regions of the country, such as Bogotá, Antioquia (Medellín), and Valle del Cauca (Cali), where people commonly migrate to get job opportunities, present a greater inequality in terms of labor condition.
Finally, we evidence an estimated effect greater than 5 points in some cases and up to 10 points in others, on math scores, for those who did not work in the previous week to perform the test. In particular, in Antioquia, Magdalena, and Santander, not working increases the math score considerably.

\section{Discussion}\label{sec:discusion}

Our findings reveal that implementing the Multinomial-Dirichlet Model works well in scenarios requiring estimating proportions of interest from surveys. Specifically, in the context of political polls, we show that the majority (73.3\%) of the credibility intervals include the observed observed vales after Election Day. On the other hand, implementing the Poisson regression model exemplifies the use of Monte Carlo simulation in scenarios where the researcher has small sample sizes to assess the relationship between variables. In particular, the Bayesian model reasonably fits the data set in population dynamics. Likewise, the operationalization of the linear regression model from a Bayesian point of view allows us to illustrate the usefulness of hierarchical modeling to characterize population groups. Specifically, in the context of the performance of standardized tests, the model makes it possible to identify regions of Colombia with outstanding scores in mathematics, aside to quantify the association that covariates such as gender and employment status have on the me math score by geographic area.

In addition, from the results in the applied contexts, we discuss and provide the technical details about conjugate modeling, hierarchical modeling, Monte Carlo simulation, Gibbs sampler,
the Metropolis-Hastings algorithm, the Monte Carlo Hamiltonian algorithm, the evaluation of the model's goodness of fit through test statistics, and the use of information criteria for model comparisson.

On the other hand, the reader must be aware of the free-use specialized software alternatives currently available for doing Bayesian computing. However, we do not discuss them in this document for space reasons. 
These include Bugs (Bayesian inference Using Gibbs Sampling), Jags (Just Another Gibbs Sampler), Stan and Nimble (e.g., \citealt{kruschke2014doing}, and \citealt{mcelreath2020statistical}), which are available in both R and Python. 
Finally, we encourage readers to inquire about other important topics typical of the Bayesian paradigm. These include interchangeability and De Finetti's representation theorem, improper priors, objective priors, Bayes factors, model averaging, approximations of the posterior distribution through analytic methods (e.g., variational inference), and Bayesian non-parametric statistics. All of these topics can be found at \cite{gelman2013bayesian}, \cite{reich2019bayesian}, and \cite{heard2021introduction}.

%\nocite{*}
\bibliographystyle{apalike}
\bibliography{References}

\appendix

\section{Multinomial-Dirichlet Model}\label{app_multinomial_dirichlet}
Let $k$ be independent random variables $X_1,\ldots, X_k$ such that $X_j\mid\alpha_i,\beta \simind \textsf{Gamma}(\alpha_j, \beta)$, for $j=1,\ldots ,k$. It can be shown that the random vector
$$
\boldsymbol{Y}=(Y_1,\ldots,Y_k)=\left(\frac{X_1}{X_1+\ldots+X_k},\ldots,\frac{X_k}{X_1+\ldots+X_k}\right)
$$
has a Dirichlet distribution with parameter $\bs{\alpha} = (\alpha_1,\ldots,\alpha_k)$, i.e., $\boldsymbol{Y}\mid\bs{\alpha} \sim\textsf{Dir}(\bs{\alpha})$.
This result leads to the following algorithm to generate random vectors $\bm{ \theta } = ( \theta_1, \dots, \theta_k )$ with Dirichlet distribution with parameter $\bs{\alpha}$:

\begin{enumerate}
    \item Choose any value for $\beta > 0$ (e.g., $\beta = 1$).
    \item Simulate $g_1,\ldots,g_k$ such that $g_j\simind\textsf{Gamma} ( \alpha_j , \beta)$, for $j=1,\ldots,k$.
    \item Compute $\theta_j = g_j/ \sum_{ \ell = 1 }^k g_\ell$, for $j=1,\ldots,k$.
\end{enumerate}

\section{Poisson regression}\label{app_regresion_poisson}

Let $\boldsymbol{\beta}^{(b)}$ be the state of parameter $\boldsymbol{\beta}$ at iteration $b$ of the algorithm, for $b = 1,\ldots,B$. Given an initial value $\boldsymbol{\beta}^{(0)}$, the following algorithms
generate a new state $\boldsymbol{\beta}^{(b)}$ from the preceding state $\boldsymbol{\beta}^{(b-1)}$.

\textbf{Metropolis Algorithm}

\begin{enumerate}
    \item Simulate $\boldsymbol{\beta}^*\sim\textsf{N}(\boldsymbol{\beta}^{(b-1)},\mathbf{\Delta}_0)$, with $\mathbf{\Delta}_0$ a fixed covariance matrix. 
    \item Compute the transition probability
    $$\alpha = \min\left\{ \exp{\left( \log p(\boldsymbol{\beta}^*\mid\boldsymbol{y}) - \log p(\boldsymbol{\beta}^{(b-1)}\mid\boldsymbol{y})   \right)}, 1 \right\}\,,$$ 
    where $\log p(\boldsymbol{\beta}\mid\boldsymbol{y})$ is given in \eqref{eq_log_beta_poisson}.
    
    \item Set 
    $$
    \boldsymbol{\beta}^{(b)} = 
    \begin{cases}
	\boldsymbol{\beta}^*,       & \text{with probability $\alpha$;}\\
    \boldsymbol{\beta}^{(b-1)}, & \text{with probability $1-\alpha$.}
	\end{cases}
    $$
\end{enumerate}

\textbf{Hamiltonian Monte Carlo algorithm}

\begin{enumerate}
    \item Simulate $\boldsymbol{\varphi}\sim\textsf{N}(\boldsymbol{0},\mathbf{M})$, with $\mathbf{M}$ a fixed covariance matrix.
    \item Establish $\boldsymbol{\beta}^* = \boldsymbol{\beta}^{(b-1)}$ and $\boldsymbol{\varphi}^* = \boldsymbol{\varphi}$.
    \item Update $(\boldsymbol{\beta}^*,\boldsymbol{\varphi}^*)$ through $L$ jumps scaled by a factor $\epsilon$ as follows:
     \begin{enumerate}
        \item Update $\boldsymbol{\varphi}^*$:
        $$\boldsymbol{\varphi}^* \leftarrow \boldsymbol{\varphi}^* - \frac{\epsilon }{2}\,\frac{\partial}{\partial \boldsymbol{\beta}}\log p(\boldsymbol{\beta}^* \mid \yv)\,,$$
        where $\frac{\partial}{\partial \boldsymbol{\beta}}\log p(\boldsymbol{\beta} \mid \yv)$ is given in  \eqref{eq_log_beta_gradiente_poisson}.
        \item Update $\boldsymbol{\beta}^*$:
        $$\boldsymbol{\beta}^* \leftarrow \boldsymbol{\beta}^* + \epsilon\, \mathbf{M}\boldsymbol{\varphi}^*$$
        \item Repeat the steps (a) y (b) $L-1$ times.
    \end{enumerate}
    \item  Compute the transition probability 
    $$\alpha = \min\left\{ \exp{\left( \log p(\boldsymbol{\beta}^*\mid\boldsymbol{y}) - \log p(\boldsymbol{\beta}^{(b-1)}\mid\boldsymbol{y}) + \log p(\boldsymbol{\varphi}^*) - \log p(\boldsymbol{\varphi})   \right)}, 1 \right\}\,,$$ 
    where $\log p(\boldsymbol{\beta}\mid\boldsymbol{y})$ is given in \eqref{eq_log_beta_poisson} y $ p(\boldsymbol{\varphi}) = \textsf{N}(\boldsymbol{\varphi}\mid\boldsymbol{0},\mathbf{M})$.
    \item Set 
    $$
    \boldsymbol{\beta}^{(b)} = 
    \begin{cases}
	\boldsymbol{\beta}^*,       & \text{with probability $\alpha$;}\\
    \boldsymbol{\beta}^{(b-1)}, & \text{with probability $1-\alpha$.}
	\end{cases}
    $$
\end{enumerate}

\section{Multiple linear regression}\label{app_regresion_lineal_multiple}

Let $\mathbf{\Theta}^{(b)}$ be the state of parameter $\mathbf{\Theta}$ at iteration $b$ of the algorithm, for $b = 1,\ldots,B$. Given an initial value $\mathbf{\Theta}^{(0)}$, the following algorithms
generate a new state $\mathbf{\Theta}^{(b)}$ from the preceding state $\mathbf{\Theta}^{(b-1)}$, by iteratively sampling the elements of $\mathbf{\Theta}$ from the corresponding complete conditional distributions.
These distributions are obtained directly from the posterior distribution of $\mathbf{\Theta}$, taking into account only the expressions that involve the component of $\mathbf{\Theta}$ we are interested in since the other terms can be regarded as constant.

\textbf{Model 1: Multiple linear regression}

Parameters: $\mathbf{\Theta} = (\bev, \sig^2)$.

Posterior distribution: 
$$
p(\mathbf{\Theta}\mid\yv)\propto \normal_n(\yv\mid\Xm\bev,\sig^2\I_n)\times \normal_p (\bev\mid\bev_0, \SIG_0) \times \GI \left(\sigma^2\mid\tfrac{\nu_0}{2},\tfrac{\nu_0\sigma^2_0}{2}\right)\,.
$$

Algorithm:
\begin{enumerate}
\item Sample $\bev$ from its full conditional distribution,
$$
\bev\mid - \sim
\textsf{N}_p\left( (\mathbf{\Sigma}_0^{-1} + \tfrac{1}{\sigma^2}\mathbf{X}^{\textsf{T}}\mathbf{X} )^{-1} (\mathbf{\Sigma}_0^{-1}\boldsymbol{\beta}_0 + \tfrac{1}{\sigma^2}\mathbf{X}^{\textsf{T}}\boldsymbol{y}  )  , (\mathbf{\Sigma}_0^{-1} + \tfrac{1}{\sigma^2}\mathbf{X}^{\textsf{T}}\mathbf{X} )^{-1} \right)\,.
$$
\item Sample $\sig^2$ from its full conditional distribution,
$$
\sig^2\mid - \sim
\textsf{GI}\left( \frac{\nu_0 + n}{2}, \frac{\nu_0\sigma^2_0 + (\yv-\Xm\bev)^{\trans} (\yv-\Xm\bev)  }{2} \right)\,.
$$
\end{enumerate}

\textbf{Model 2: Multiple linear regression with random effects}

Parameters: $\mathbf{\Theta} = (\theta_1,\ldots,\theta_m,\tau^2, \bev, \sig^2)$.

Posterior distribution:
\begin{align*}
p(\mathbf{\Theta}\mid\yv) &\propto \normal_n(\yv\mid\Xm\bev + \bs{\vartheta},\sig^2\I_{n}) \\
&\qquad\times \prod_{j=1}^m \normal(\theta_j\mid 0,\tau^2) \times \GI\left(\tau^2\mid\tfrac{\eta_0}{2},\tfrac{\eta_0\tau^2_0}{2}\right)\\
&\qquad\qquad\times \normal_p (\bev\mid\bev_0, \SIG_0) \times \GI \left(\sigma^2\mid\tfrac{\nu_0}{2},\tfrac{\nu_0\sigma^2_0}{2}\right)\,.
\end{align*}

Algorithm:
\begin{enumerate}
\item Sample $\theta_j$, $j=1,\ldots,m$, from its full conditional distribution,
$$
\theta_j\mid - \sim
\normal\left( \frac{ \frac{1}{\sig^2} \sum_{i=1}^{n_j} (y_{i,j} - \xv_{i,j}^{\trans}\bev) }{\frac{1}{\tau^2} + \frac{n_j}{\sig^2}} , \frac{1}{\frac{1}{\tau^2} + \frac{n_j}{\sig^2}} \right)\,.
$$
\item Sample $\tau^2$ from its full conditional distribution,
$$
\tau^2\mid - \sim
\GI\left(\frac{\eta_0+m}{2} , \frac{\eta_0\tau_0^2 + \tev^{\trans} \tev}{2} \right)\,.
$$
\item Sample $\bev$ from its full conditional distribution,
$$
\bev\mid - \sim
\normal_p\left( \left(\SIG_0^{-1} + \tfrac{1}{\sigma^2}\Xm^{\trans}\Xm \right)^{-1} \left( \SIG_0^{-1}\bev_0 + \tfrac{1}{\sigma^2}\Xm^{\trans}(\yv-\bs{\vartheta}) \right) , \left(\SIG_0^{-1} + \tfrac{1}{\sigma^2}\Xm^{\trans}\Xm \right)^{-1} \right)\,.
$$
\item Sample $\sig^2$ from its full conditional distribution,
$$
\sig^2\mid - \sim
\textsf{GI}\left( \frac{\nu_0 + n}{2}, \frac{\nu_0\sigma^2_0 + (\yv-\Xm\bev-\bs{\vartheta})^{\trans} (\yv-\Xm\bev - \bs{\vartheta})}{2} \right)\,.
$$
\end{enumerate}

\textbf{Model 3: Multilevel multiple linear regression with random effects}

Parameters:  $\mathbf{\Theta} = (\theta_1,\ldots,\theta_m,\tau^2, \bev_1,\ldots,\bev_m, \bev, \Sig, \sig^2_1,\ldots,\sigma^2_m, \nu, \sig^2)$.

Posterior distribution:
\begin{align*}
p(\mathbf{\Theta}\mid\yv) &\propto  \prod_{j=1}^m\normal_{n_j}(\yv_j\mid\Xm_j\bev_j + \theta_j\bs{1}_{n_j},\sig_j^2\I_{n_j}) \\
&\qquad\times \prod_{j=1}^m \normal(\theta_j\mid 0,\tau^2) \times \GI\left(\tau^2\mid\tfrac{\eta_0}{2},\tfrac{\eta_0\tau^2_0}{2}\right)\\
&\qquad\qquad\times \prod_{j=1}^m \normal_p (\bev_j\mid\bev, \SIG) \times \normal_n(\bev\mid\bs{\mu}_0,\mathbf{\Lambda}_0) \times  \textsf{WI}(\SIG\mid n_0, \Sm^{-1}_0)\\
&\qquad\qquad\qquad\times \prod_{j=1}^m \GI \left(\sig^2_j\mid \tfrac{\nu}{2}, \tfrac{\nu\sig^2}{2} \right) \times e^{-\kappa_0\nu} \times \textsf{G}(\sigma^2\mid\alpha_0, \beta_0)
\,.
\end{align*}

Algorithm:
\begin{enumerate}
\item Sample $\theta_j$, $j=1,\ldots,m$, from its full conditional distribution,
$$
\theta_j\mid - \sim
\normal\left( \frac{ \frac{1}{\sig^2} \sum_{i=1}^{n_j} (y_{i,j} - \xv_{i,j}^{\trans}\bev_j) }{\frac{1}{\tau^2} + \frac{n_j}{\sig^2}} , \frac{1}{\frac{1}{\tau^2} + \frac{n_j}{\sig^2}} \right)\,.
$$
\item Sample $\tau^2$ from its full conditional distribution,
$$
\tau^2\mid - \sim
\textsf{GI}\left(\frac{\eta_0+m}{2} , \frac{\eta_0\tau_0^2 + \tev^{\trans} \tev}{2} \right)\,.
$$
\item Sample $\bev_j$, $j=1,\ldots,m$, from its full conditional distribution,
$$
\bev_j\mid - \sim
\normal_p\left( \left(\SIG^{-1} + \tfrac{1}{\sigma_j^2}\Xm_j^{\trans}\Xm_j \right)^{-1} \left( \SIG^{-1}\bev + \tfrac{1}{\sigma_j^2}\Xm_j^{\trans}(\yv_j-\theta_j\bs{1}_{n_j}) \right) , \left(\SIG^{-1} + \tfrac{1}{\sigma^2_j}\Xm_j^{\trans}\Xm_j \right)^{-1} \right)\,.
$$
\item Sample $\bev$ from its full conditional distribution,
$$
\bev\mid - \sim
\normal_p\left( \left(\LAM_0^{-1} + m\SIG^{-1} \right)^{-1} \left( \LAM_0^{-1}\muv_0 + \SIG^{-1}\textstyle\sum_{j=1}^m \bev_j \right) , \left(\LAM_0^{-1} + m\SIG^{-1} \right)^{-1} \right)\,.
$$
\item Sample $\Sig$ from its full conditional distribution,
$$
\Sig\mid - \sim
\textsf{WI}\left(n_0 + m, \left( \Sm_0 + \textstyle\sum_{j=1}^m (\bev_j - \bev)(\bev_j - \bev)^{\trans} \right)^{-1} \right)\,.
$$
\item Sample $\sig_j^2$, $j=1,\ldots,m$, from its full conditional distribution,
$$
\sig_j^2\mid - \sim
\GI\left(\frac{\nu+n_j}{2} , \frac{\nu\sigma^2 + (\yv_j - \Xm_j\bev_j - \theta_j\bs{1}_{n_j})^{\trans} (\yv_j - \Xm_j\bev_j - \theta_j\bs{1}_{n_j})}{2} \right)\,.
$$
\item Sample $\nu$ from its full conditional distribution,
$$
p\left(\nu \mid - \right)
\propto\left[\frac{\left(\nu \sigma^{2} / 2\right)^{\nu / 2}}{\Gamma\left(\nu / 2\right)}\right]^{m}
\left[\prod_{j=1}^m \sigma_j^{-2}\right]^{\nu / 2}
\expo{ -\nu\left(\kappa_0+\frac{\sigma^{2}}{2} \sum_{j=1}^m \sigma_{j}^{-2}\right) }\,.
$$
\item Sample $\sig^2$ from its full conditional distribution,
$$
\sig^2\mid - \sim
\textsf{G}\left( \alpha_0+\frac{m \nu}{2}, \beta_0+\frac{\nu}{2} \sum_{j=1}^m \sigma_{j}^{-2}\right)\,.
$$
\end{enumerate}

\section{Notation}

The Gamma function is denoted by $\Gamma(\cdot)$ and is given by $\Gamma(x)=\int_0^\infty u^{x-1}\,e^{-u}\,\text{ d}u$.
Matrices and vectors with entries consisting of subscripted variables are denoted by the variable letter in bold.
For example, $\xv = (x_1,\ldots, x_n)$ denotes a column vector of $n\times1$ with entries $x_1,\ldots, x_n$.
We use $\boldsymbol{0}$ and $\boldsymbol{1}$ to denote the column vector whose entries are equal to 0 and 1, respectively, and we also use $\I$ to denote the identity matrix.
A subscript in this context indicates the corresponding dimension. For example, $\I_n$ denotes the identity matrix of size $n\times n$.
The transpose of a vector $\xv$ is denoted by $\xv^\textsf{T}$. Similarly for matrices.
Also, if $\Xm$ is a square matrix, we use $\text{tr}(\Xm)$ and $|\Xm|$ to denote the trace and determinant of $\Xm$, respectively.

Below we present the probabilistic distributions used in the applications:
\begin{itemize}

        \item Gamma:

        A random variable $X$ has a Gamma distribution with parameters $\alpha$ and $\beta$, denoted by $X\mid\al,\be\sim\textsf{G}(\al,\be)$, if the probability density function is
        $$
        p(x\mid\al,\be) = \frac{\be^\al}{\Gamma(\al)}\,x^{\al-1}\,\exp{\{-\be x \}}\,,\quad x>0\,,\quad \alpha>0\,,\quad \beta>0\,.
        $$

	\item Inverse Gamma:

        A random variable $X$ has an inverse gamma distribution with parameters $\alpha$ and $\beta$, denoted by $X\mid\al,\be\sim\textsf{GI}(\al,\be)$ , if the probability density function is
        $$
        p(x\mid\al,\be) = \frac{\be^\al}{\Gamma(\al)}\,x^{-(\al+1)}\,\exp{\{- \be/x \}}\,,\quad x>0\,,\quad \alpha>0\,,\quad \beta>0\,.
        $$

        \item Normal:

        A random variable $X$ has a Normal distribution with parameters $\mu$ and $\sig^2$, denoted by $X\mid\mu,\sig^2\sim\textsf{N}(\mu,\ sig^2)$, if the probability density function is
        $$
        p(x\mid\mu,\sig^2) = \frac{1}{\sqrt{2\pi\sig^2}}\,\exp{\left\{-\frac{1}{2} \, \frac{(x-\mu)^2}{\sigma^2} \right\}}\,,\quad x\in\mathbb{R}\,,\quad \mu\in\mathbb{ R}\,,\quad \sig^2 > 0\,.
        $$

    \item Dirichlet: 

        A random vector $\Xv = (X_1,\ldots, X_K)$ has a Dirichlet distribution with parameter $\bs{\alpha}$, denoted by $\Xv\mid\bs{\alpha}\sim\textsf{Dir}(\bs{\alpha})$, if the probability density function is
        $$
        p(x\mid\bs{\alpha}) =
        \left\{
        \begin{array}{ll}
        \frac{\Gamma\left(\sum_{k=1}^K\al_k\right)}{\prod_{k=1}^K\Gamma(\al_k)}\prod_{k=1}^K x_k ^{\al_k-1}, & \hbox{if $\sum_{k=1}^K x_k = 1$, $\al_1,\ldots,\alpha_K > 0$;} \\
        0, & \hbox{otherwise.}
        \end{array}
        \right.
        $$

	\item Multivariate Normal:

        A $d\times 1$ random vector $\Xv=(X_1\ldots,X_d)$ has a Multivariate Normal distribution with parameters $\muv$ and $\Sig$, denoted by $\Xv\mid\muv,\Sig \sim \textsf{N}_d(\muv,\Sig)$, if the probability density function is
        $$
        p(\xv\mid\muv,\Sig) = (2\pi)^{-d/2}\,|\Sig|^{-1/2}\,\exp{\left\{-\tfrac12 (\xv - \muv)^{\textsf{T}}\Sig^{-1}(\xv - \muv) \right\}}\,,\quad \xv\in\mathbb{R}^d \,,\quad \muv\in\mathbb{R}^d\,,\quad \Sig>0\,.
        $$

	\item Inverse Wishart:

        A $d\times d$ random matrix $\mathbf{W}$ has an Inverse Wishart distribution with parameters $\nu$ and $\Sm^{-1}$, denoted by $\mathbf{W} \sim \textsf{WI}(\nu, \Sm^{-1})$, if the probability density function is
        $$
        p(\mathbf{W}) \propto |\mathbf{W}|^{-(\nu+d+1)/2}\,\exp{\left\{-\tfrac12\text{tr}(\Sm\mathbf{W}^{-1}) \right\}}\,,\quad \mathbf{W}>0\,,\quad \nu>0\,,\quad \Sm > 0.
        $$

\end{itemize}

\end{document}